\documentclass{aa}  
\usepackage{graphicx}
\usepackage{txfonts}
\usepackage{xcolor}
\usepackage[T1]{fontenc}
\usepackage{ae,aecompl}
\usepackage{microtype}
\usepackage{epsfig}
\usepackage{amsmath}
\usepackage{natbib}
\usepackage{bm}
\usepackage[]{hyperref}
\usepackage{subfigure}
\usepackage[normalem]{ulem}
\usepackage{textcomp}
\newcommand{\Fig}[1]{{Fig.\ \ref{fig:#1}}}
\newcommand{\Figure}[1]{{Figure \ref{fig:#1}}}

\newcommand{\Eq}[1]{Eq.\ \ref{eq:#1}}

\usepackage{orcidlink}

\makeatletter
\renewcommand*\aa@pageof{, page \thepage{} of \pageref*{LastPage}}
\makeatother

\begin{document}

   \title{Global MHD simulations of the solar convective zone using a volleyball mesh decomposition. I. Pilot}
   
   \titlerunning{Volleyball Sun}

   \author{A. Popovas
          \inst{1,2}\orcidlink{0000-0003-4436-0764},
          {\AA}. Nordlund\inst{1,3}\orcidlink{0000-0002-2219-0541},
          M. Szydlarski \inst{1,2}\orcidlink{0000-0002-9115-4448}
          \and
          Y. Zhou\inst{1,2}\orcidlink{0000-0003-0817-6126}
          }

   \institute{Rosseland Centre for Solar Physics, University of Oslo, P.O. Box 1029, Blindern, NO-0315 Oslo, Norway\\
\email{andrius.popovas@astro.uio.no}
\and
Institute of Theoretical Astrophysics, University of Oslo, P.O. Box 1029, Blindern, NO-0315 Oslo, Norway
\and
Niels Bohr Institute, University of Copenhagen, \O ster Voldgade 5-7, DK-1350 Copenhagen, Denmark
}

\date{Received; accepted}

 
  \abstract
   {Solar modelling has long been split into `internal' and `surface' modelling because of the lack of tools to connect the very different scales in space and time, as well as the widely different environments and the dominating physical effects involved. Significant efforts have recently been put into resolving this disconnect.}
   {We address the outstanding bottlenecks in connecting internal convection zone and dynamo simulations to the surface of the Sun, and conduct a proof-of-concept high-resolution global simulation of the convection zone of the Sun, using the task-based DISPATCH code framework.}
   {We present a new `volleyball' mesh decomposition, which has Cartesian patches tessellated on a sphere with no singularities. We use our new entropy based HLLS approximate Riemann solver to model magneto-hydrodynamics in a global simulation, ranging between 0.655---0.998 R$_\odot$, with an initial ambient magnetic field set to 0.1 Gauss. This paper describes the methods and the pilot runs of the setup.}
   {The mesh decomposition exhibits neither unwanted artefacts, nor noise build-up. The simulations develop convective motions with complex, turbulent structures. Small-scale dynamo action twists the ambient magnetic field and locally amplifies magnetic field magnitudes by more than two orders of magnitude within the initial run-time.}
   {}


   \maketitle
   \nolinenumbers
\section{Introduction}

Throughout most of the history of trying to understand the Sun, solar research has been split into `internal' and `surface' solar physics topics, and an integrated global view of the Sun's complex plasma dynamics has been lacking. For instance, surface models, however complex they have become (e.g.\ \citet{Carlsson_2016,Chen_2022,Przybylski_2022}), have to make certain initial and boundary assumptions about the magnetic field. These assumptions are usually specified as adjustable input parameters, reflecting the fact that  larger scale phenomena beyond the simulation domain, such as differential rotation and magnetic fields, affect the dynamics within. Moreover, local simulations are usually plane-parallel, which severely restricts the horizontal extent that can be simulated without sacrificing realism because of lacking curvature and large scale differential rotation. In addition, certain features in the atmosphere of the Sun can span large distances, e.g.\ coronal loops $\sim$ 100 Mm \citep{Reale_2010}, prominences $\sim$ 10--100 Mm \citep{Parenti_2014}, and supergranulation $\sim$ 30--40 Mm \citep{Nordlund_2009,Rieutord_2010}. Such features are thus difficult to study in isolation.

Likewise, global convection and dynamo simulations, seeking to reveal the origin of the magnetic field and its cyclic behaviour, typically use spherical geometry and usually span from the bottom of the convection zone up to 10--30 Mm below the surface, which gives significant benefits (large time-step, use of incompressible gas approximations, etc.). They can reproduce differential rotation and large-scale transport, but they exclude the near-surface layers where compressibility, radiative losses, and the strongest stratification shape the convective driving at the cost of losing the magnetic field's observable manifestations, e.g. the existence of sunspots, and the ability to make direct comparison with observations. \cite{Miesch_2008} ran global anelastic simulations of the solar convection zone up to 0.98 $\rm{R}_\odot$ (around 15 Mm below the surface) revealing that coherent downflow structures associated with giant cells play a significant role in maintaining the differential rotation. \cite{Hotta_2014,Hotta_2015} ran global convection simulations of non-rotating and rotating solar models. They used a `reduced speed of sound' technique (RSST, e.g. \citealt{Hotta_2012}), spherical geometry and a Yin-Yang grid \citep{Kageyama_2003}, with the radial extent reaching up to 0.99 $\rm{R}_\odot$. The work was expanded by \cite{Hotta_2021,Hotta_2022} to study solar differential rotation, spanning 0.71--0.96 $\rm{R}_\odot$. \cite{Guerrero_2022} ran global anelastic convection simulations between 0.6--0.96 $\rm{R}_\odot$ and found that obtaining the correct distribution of angular momentum is not a mere issue of numerical resolution; magnetism and the presence of a near-surface shear layer may be necessary to simulate the solar interior accurately\citep{Guerrero_2022}. 

To resolve this disconnect many bottlenecks must be addressed: Highly disparate spatial and temporal scales, complex microphysics, the interplay between local and global effects, very wide dynamic range, and required resolution in both space and time to resolve various features at different depths of the atmosphere. Tremendous efforts have been made recently to resolve such disconnects for the Sun and other stars.

Extended Cartesian boxes were used to take the first steps towards fully connecting the deep convective interiors of Sun and stars to their photospheres. For example, \cite{Stein_2012} used a 48$\times$48$\times$20 Mm box to simulate an active region without any initial flux structures. Nevertheless, a small active region was formed from a uniform, untwisted, horizontal magnetic field of 1 kilogauss advected into the computation domain at the bottom boundary. \cite{Hotta_2019} ran a hydrodynamic simulation of the full vertical extent of convection zone, with a 240$\times$240 Mm horizontal extent. The simulation did not have the effects of rotation included. The authors found that the surface region has an unexpectedly weak influence on the deep convection zone. This work was then continued by \cite{Hotta_2020} where the authors ran a flux emergence simulation in a very large (98.3$\times$98.3$\times$200 Mm) box, vertically covering the entire convection zone and extending into the photosphere. The authors found indications that a deep simulation domain is needed for flux emergence simulations. 

In a larger context, the Sun is only one example of a rather common type of stars, and similar efforts have been made for other stars; e.g.\ \cite{Kapyla_2021} ran 288$^3$ and 576$^3$ star-in-a-box simulations of fully convective stars, showing that there are large similarities between partially and fully convective stars when it comes to generating differential rotation and large-scale magnetism. \cite{Freytag_2017} used the CO5BOLD radiation-hydrodynamics code to produce a grid of star-in-a-box models of the outer convective envelope and the inner atmosphere of AGB stars;  
Global radiation-hydrodynamic simulations for the convective envelope of red supergiants were carried out by \citet{Goldberg_2022} with the \texttt{Athena$++$} code, in a spherical-polar mesh.
\cite{Brun_2017} ran hydrodynamic simulations of 15 stars with variable mass and rotation period ($\Omega_\star$) to study how the change of global rotation rate influences the overall dynamics and large scale flows arising in the convective envelopes of stars all the way up to  0.97 of stellar radius. Later \cite{Brun_2022} expanded these simulations by introducing magnetic fields and found, as expected, differences in resulting differential rotation profiles, which vary with $\Omega_\star$. Changes of $\Omega_\star$ also lead to transitions in the nature of the dynamo processes (short or longer duration cycles or no cycles).

The aforementioned works are just a few examples, using rather different mesh decomposition, physics involved, spatial and time resolution, etc.. What they have in common is an effort to connect the deep interiors with the atmospheres of stars. Such simulations are extremely computationally expensive, often forcing significant sacrifices, in numerical resolution, radial extent, and/or realism of the equation-of-state. Incompressible or anelastic models are not applicable close to the photosphere of stars, Cartesian boxes are not very suitable for maintaining accurate spherical hydrostatic equilibrium, and spherical coordinates have singularities at the poles, to mention just a few limitations. 

In this work we develop and demonstrate a discretisation designed explicitly for this regime: a ``volleyball'' mesh decomposition, in which the sphere is covered by many small, locally Cartesian patches tessellated on a six-face ``volleyball'' topology. Each patch is evolved using a standard Cartesian Godunov-type magnetohydrodynamic (MHD) method, while geometric effects (e.g. gravity direction) are applied consistently in the local patch frames. The absence of coordinate singularities and the locality of patch communication make this approach well suited to task-based execution, local time stepping, and (static or adaptive) mesh refinement. The main technical concern in this patch-based spherical decomposition is whether inter-patch exchange -- particularly across the face seams -- introduces numerical artefacts, spurious noise, or long-term drift in conserved quantities and in $\nabla \cdot B$. We describe and validate the inter-patch coupling machinery -- region-of-interest (ROI) and region-of-authority (ROA) mapping and guard-zone exchange -- for both cell-centred and face-centred MHD variables, including the transformations required between patch frames.

Due to page limitations, this paper focuses on the method itself and demonstrates that the coupled patch system is stable and free of seam-driven defects in a global, stratified MHD convection setup. It establishes the numerical and algorithmic basis required for subsequent studies that will extend the physics (e.g. radiative transfer in the surface layers, demonstrated in Appendix \ref{sec:photosphere}) and increase resolution. Therefore, the present paper does not aim to deliver converged solar-science results.

Our pilot implementation is carried out in the DISPATCH framework \citep{Nordlund_2018MNRAS}, which is designed for large-scale task-based execution on distributed-memory systems. DISPATCH enables local time stepping per patch and restricts MPI communication to nearest neighbours, providing a practical path to extreme-scale simulations. 

We demonstrate with this global pilot run that the method described in this paper sustains a stable stratified convection state without visible seam artefacts and that magnetic fields can be evolved robustly, including early small-scale dynamo amplification from a weak initial field.

The paper is organised as follows. Section \ref{sec:numerical_scheme} introduces the numerical scheme used for MHD evolution. Section \ref{sec:domain_decomposition} presents the volleyball mesh decomposition and the coupling between patches. Section \ref{sec:experiment_design} describes the pilot experiment design, including boundary conditions and the near-surface driving used in this initial stage. Section \ref{sec:relax} presents verification and pilot-run diagnostics emphasising seam behaviour, equilibrium maintenance, and early MHD evolution. Section \ref{sec:discussion} discusses limitations of the pilot setup and outlines the next steps toward fully realistic high-resolution global simulations including the surface. 
\section{Numerical scheme}
\label{sec:numerical_scheme}
In this pilot study we solve the equations of ideal compressible MHD in a gravitationally stratified, rotating frame, supplemented by parametrised heating/cooling terms used to drive and relax the convection state. The numerical method is a finite-volume, second-order Godunov scheme, using an entropy-based approximate Riemann solver (HLLS) and constrained transport (CT) for the induction equation, see \cite{Popovas_2025a} for full details. This solver is advantageous in highly stratified flows where maintaining a near-hydrostatic equilibrium and avoiding thermodynamic pathologies is essential. The mesh decomposition consists of locally Cartesian patches tessellated onto a sphere with the gravity vector pointing towards the centre and appropriate rotation terms (Sections \ref{sec:domain_decomposition}–\ref{sec:experiment_design}). 

All non-conservative terms (gravity, rotation, heating/cooling) are applied as source terms in both predictor and corrector steps, consistent with the temporal discretisation of the scheme. The fluid equations for the total mass density, momentum density, entropy per unit mass, and the magnetic field of the system can be written as a system of conservation laws,

\begin{equation}
    \frac{\partial}{\partial t} \left[ \begin{array} {c} \rho \\ \rho \boldsymbol{u} \\ \rho S \\ \boldsymbol{B} \end{array} \right] +
    \nabla \cdot \left[ \begin{array} {c} \rho \boldsymbol{u} \\ \rho (\boldsymbol{u} \otimes \boldsymbol{u}) + P_{tot} - \boldsymbol{B} \otimes \boldsymbol{B}   \\ \rho \boldsymbol{u} S \\ \boldsymbol{B} \otimes  \boldsymbol{u} - \boldsymbol{u} \otimes \boldsymbol{B} \end{array} \right] = \left[ \begin{array} {c} 0 \\ \boldsymbol{\Phi}\\ \boldsymbol{Q}/T_{gas} + \boldsymbol{S_{g}} \\ 0 \end{array} \right],
\label{eq:basic_eq_set_S}
\end{equation}
with
\begin{equation}
\label{eq:divb}
    \nabla \cdot \boldsymbol{B} = 0,
\end{equation}
where $\rho$, $\boldsymbol{u}$, $S$, $P_{tot}$, and $\boldsymbol{B}$ are density, velocity, entropy per unit mass, total pressure and magnetic field respectively; $\Phi$ is force per unit volume, which includes force of gravity, Coriolis force, etc; $\boldsymbol{Q}$ is heating per unit volume, which includes Newton cooling, radiative heat transfer, etc; $T_{gas}$ is gas temperature and $\boldsymbol{S_{g}}$ is entropy generation from converting kinetic and magnetic energy to heat. The system is closed by an equation of state (EOS), described below.

The rotation and gravity vectors are evaluated in the local patch frames; the coupling of patches on the sphere is described in Section \ref{sec:domain_decomposition}.

When patch-to-patch guard-zone values are interpolated in space and time, the resulting magnetic field in the guard zones is not guaranteed to be strictly divergence-free. We therefore apply a local guard-zone divergence-cleaning correction and, in addition, an elliptic (Poisson) divergence-cleaning step based on the Poisson solver of \cite{Ramsey_2018}, which mitigates any residual $\nabla \cdot \mathbf{B}$ introduced by inter-patch interpolation and round-off noise.
\subsection{Conservation properties of the scheme}
The numerical scheme conserves  the total mass density, momentum density  and the magnetic field inside patches remains divergence free to numerical precision. Entropy is conserved exactly in adiabatic flows, but is by construction not conserved in the presence of shocks, and steepening of slow/fast magnetosonic waves, which can generate entropy. The conservation of the angular momentum to numerical precision can be rather challenging but it is approximately conserved by the experimental design. The details are discussed in other sections, below, but can be summarised as follows:
\begin{itemize}
    \item the finite volume methods we employ conserve fluxes exactly across cell faces within a patch, and approximately across patch boundaries;
    \item vector quantities are rotated appropriately when communicating between tilted patches;
    \item we don't employ artificial viscosity or hyperdiffusion, which could lead to spurious angular momentum transport;
    \item the `volleyball’ mesh rotates with a fixed rotation rate, and we apply the corresponding accelerations from Coriolis and centripetal forces.
\end{itemize}
Conservation between tilted patches is challenging \citep{Peng_2006,Chen_fc_2022}, and we have developed flux corrections for tilted patches with independent timestep and large stratification. Given it is a topic on its own, it is beyond the scope of this paper and will be described in detail in upcoming series of publications (Popovas et al, in prep.)
\subsection{Equation of state}
\label{sec:EOS}
We tabulate the equation of state with the FreeEOS \citep{Irwin_2012} package, adopting standard contemporary elemental abundances (e.g. \cite{Asplund_2021}). The native tables are functions of ($\rho,T$); for use with the entropy-based solver we remap them to ($\rho, S$) by computing entropy consistently from thermodynamic derivatives,
\begin{equation}
  dS = \left[ \frac{1}{T} \left( \frac{\partial \varepsilon}{\partial T}  \right)_\rho \right] dT + \left[ \frac{1}{T} \left( \frac{\partial \varepsilon}{\partial \ln\rho}  \right)_T - \frac{P}{T \rho}\right] d\ln\rho,
\end{equation}
 where $\varepsilon$ is internal energy per mass; and interpolating all required quantities onto a ($\rho, S$) grid. We verify the remapped table by checking thermodynamic identities:
 \begin{equation}
    \left( \frac{\partial \varepsilon}{\partial \rho} \right)_T = \frac{P}{\rho^2} - \frac{T}{\rho^2} \left( \frac{\partial P}{\partial T} \right)_\rho = \frac{P}{\rho^2} \left[1 - \left( \frac{\partial \ln{P}}{\partial \ln{T}}  \right)_\rho  \right],
\end{equation}
and by confirming that the ($\rho, S$) representation reproduces the original ($\rho, T$) quantities to numerical precision.
The EOS is used both to close the MHD system above and to construct the modified hydrostatic initial state described in Section \ref{sec:initial_hse}.
\subsection {Code units}
For numerical conditioning we use a rescaling with representative solar convection-zone scales (length, time, and density), and express all quantities in corresponding code units. The specific choice used in this work is provided together with the remaining pilot parameters in Section \ref{sec:experiment_design} (Table \ref{tab:sim_setup}).

\section{Domain decomposition} \label{sec:domain_decomposition}
\def\rot{\underline{\underline{R}}}
\def\src{\mathrm{source}}
\def\tgt{\mathrm{target}}
\def\half{{1/2}}

\begin{figure}
\centering
    \includegraphics[width=0.85\columnwidth]{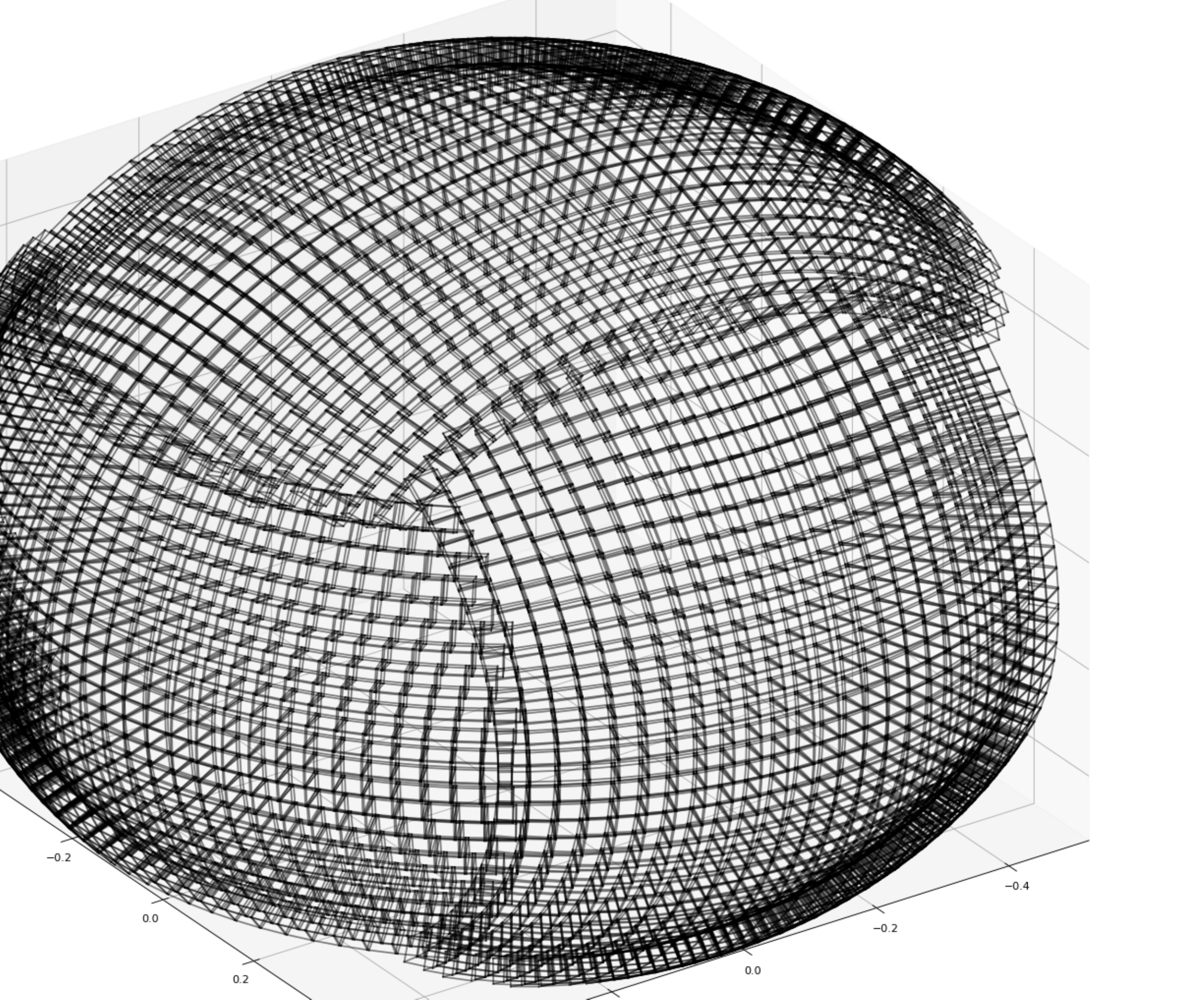}
    \caption{One layer of a volleyball mesh, with the back side removed, for clarity.}
    \label{fig:vb_mesh_3d}
\end{figure}

In DISPATCH, we use a set of Cartesian patches arranged in experiment-specific geometries. Each patch contains $N^3$ cells. In the current work, we generally use $N=24$, plus 3 guard zone cells on each side, thus 30$^3$ cells in total. These patches are then placed on a sphere, in a volleyball-like arrangement, cf.\ \Fig{vb_mesh_3d}. The patches are oriented so their $z$ axes always lie in the radial direction. The tilt between adjacent patches is very small, with exceptions at the `seams' between the six `faces' of the volleyball. 

\subsection{Volleyball patch arrangement}

This mesh decomposition may, from a first glance, appear similar to a cubed-sphere \citep{Ronchi_1996}. However, the cubed-sphere is a truly curvi-linear coordinate system in each face region, in which case the solver must be built with the metric coefficients of that coordinate system explicitly known and used. In the volleyball decomposition we have purely Cartesian patches, arranged on a sphere with  slight overlaps between them.  Note that there is no plane-parallel approximation employed; the force of gravity, for example, is everywhere pointing in the exact radial direction, and as a consequence, hydrostatic solutions inside the patches are small spherical cut-outs, which---despite their small size---deviate ever so slightly from plane-parallel solutions.

We start by tessellating a sphere by generating a set of points. The point object \texttt{P} stores as attributes
\begin{enumerate}
    \item its position both in spherical and Cartesian coordinates,
    \item the size of the patch,
    \item its north vector $\Xi$, and
    \item the coordinate system unit vectors $\rot$.
\end{enumerate}
An important value here---a horizontal step---is the angular spacing between the patches. It is defined as $\Delta \theta = \theta_0 / N_\theta$, where $\theta_0 = \tan^{-1}(2^{-\half})$ (the angular half-width of a volleyball face) and $N_\theta$ is a parameter that can vary from layer to layer (see below). We start with the outermost layer, i.e.\ $r = r_{max} = 0.998$, with half a patch size $L$ north-east offset from the centre of a front volleyball face (there are 6 faces in total---front, back, left, right, top and bottom---alternatively referred to as front, back, west, east, south and north). Given the number of intervals $N_\theta$ in the $\theta$ direction, we generate the point positions and patch sizes $L$,

\begin{equation}
    L= r \frac{2 ~ \tan{(0.5  ~ \Delta \theta)}}{1-\tan{\left(0.5 ~ \Delta \theta\right)}\left(1+2 ~ n_g/[n-1]\right)},
\end{equation}
where the last term in the denominator slightly extends the overlap to account for (and cover) the number of guard zones ($n_g =3$ in this experiment). \\
We cover the whole face by stepping $\Delta\phi$ eastwards, i.e.\ the new $\phi$ coordinate of the point is then
\begin{equation}
    \phi = \phi_{old} + \frac{\Delta\theta}{\cos(\theta) + \sin{(\theta)} ~ \Delta \theta / 2}.
\end{equation}
This stepping, interleaved with steps northward,
\begin{equation}
    \theta = \theta_{old} + \Delta\theta,
\end{equation}
ends up covering a quarter of a face (1/24 of the entire surface). The rest of the surface is covered by executing a sequence of symmetry operations---we spawn corresponding points on the other 5 volleyball faces by applying three \texttt{permutation} operations (going from front to right, to top, returning to the front) and two \texttt{reflection} operations (going from front to back). These operations have several sub-steps in them; e.g.\ for \texttt{permutation}:
\begin{enumerate}
    \item permute the Cartesian coordinates;
    \item permute a North proxy;
    \item convert Cartesian coordinates to spherical;
\end{enumerate}
and for \texttt{reflection}:
\begin{enumerate}
    \item invert Cartesian coordinates;
    \item convert Cartesian coordinates to spherical.
\end{enumerate}
After the \texttt{symmetry} operation, we go from the east to the west quadrant of the face (an \texttt{east-west} operation): 
\begin{enumerate}
    \item $\phi_{new} = -\phi$;
    \item convert spherical coordinates to Cartesian;
\end{enumerate}
In the same way, we need to go from the north to the south quadrant (a \texttt{north-south} operation): 
\begin{enumerate}
    \item $\theta_{new} = -\theta$;
    \item convert spherical coordinates to Cartesian.
\end{enumerate}
We do the last two operations twice more to get back to the original starting point and do all the symmetry operations in-between these steps.

We continue by stepping east, and keep stepping as long as the point is entirely inside the front face, i.e.\ inside the latitude limit of the front volleyball section, and as long as its lower south west patch corner is outside of the east face. When the patch goes out of bounds in the east direction we go back to the starting point and take a step north. We keep repeating the steps until the whole surface of the particular layer is covered. When this is done we are ready to move one layer down. The step in the radial direction is
\begin{equation}
\Delta r = L ~ \cos{\left( \frac{\Delta \theta}{2}\right)},
\end{equation}
where the second term is giving a slight decrease in the radial step to account for minor angles between adjacent patches in the layer above.

\subsubsection{Patch size dependency on the radius}
\begin{figure}
\centering
    \includegraphics[width=0.85\columnwidth]{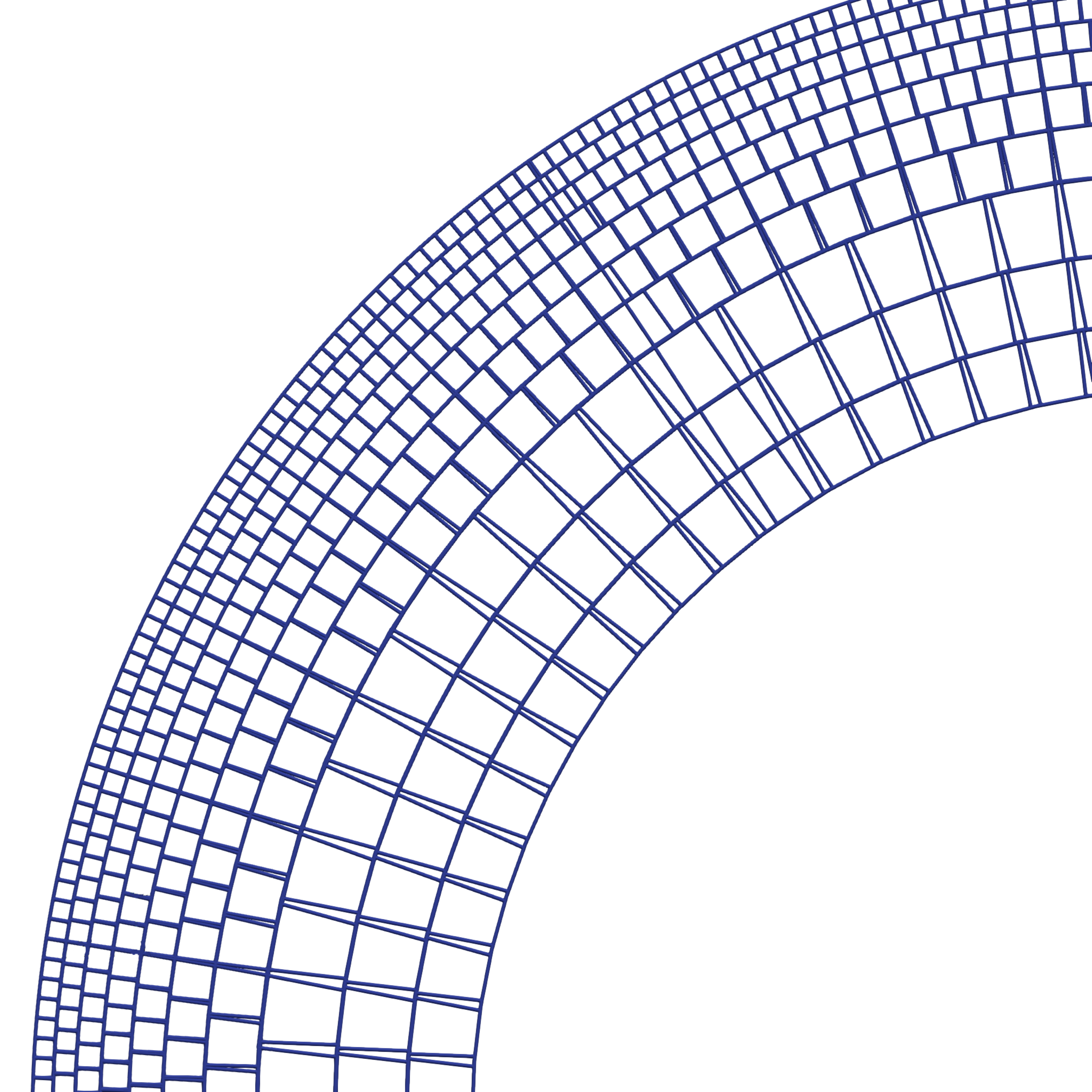}
    \caption{Radial cut of the volleyball layers.}
    \label{fig:vb_mesh_rad_slice}
\end{figure}

In spherical coordinates a volume with a constant angular size $\Delta\theta$ would become physically smaller and smaller with diminishing radius. However, in the interior of the Sun, the size of features and typical timescales are increasing with diminishing radius. Another important fact is the increase of the speed of sound with diminishing radius---together with a decreasing size this can severely restrict the allowed time-step. Instead, we want the cell size to increase with decreasing radius, to adjust to the increasing scales of motion and thus optimise the performance of the simulation, while still resolving vertical scale heights with similar number of points in all layers. To accomplish this we reduce $N_\theta$, and thus make $\Delta \theta$ larger. Extra care must be taken so that  $N_\theta$ does not become large enough for gaps to appear in the tessellation. To avoid gaps, we define a smallest allowed $N_\theta$ (typically 24 with 16 giving a marginal overlap). \Figure{vb_mesh_rad_slice} shows an example radial cut through such layers. As may be seen, patches near the surface are much smaller than patches near the bottom of the model. This alleviates the severe restriction on the time step from the CFL condition.

\subsubsection{Rotation matrices}
To rotate the patches into their correct orientation we use rotation matrices $\rot$. These matrices have the unit vectors of the local patch coordinate system expressed in the global Eulerian coordinates as columns and the unit vectors of the global coordinate system expressed in local coordinates as rows. The rotation matrix is thus:
\begin{equation}
    \rot = 
\begin{bmatrix}
\ \iota & \kappa & \lambda \ \\
\ \mu & \nu & \xi\ \\
         \tilde{x} &          \tilde{y} & \tilde{z}\ 
\end{bmatrix}
\end{equation}
where $\tilde{P} = [\tilde{x}, \tilde{y}, \tilde{z}] = P/(P \cdot P)^\half$ is the normalised Cartesian position $P$, representing unit vector `up', $\zeta = \Xi \times \tilde{P}$, $\Psi = [\iota,\kappa,\lambda]=\zeta/(\zeta \cdot \zeta)^\half$ representing a `unit vector east', $\vartheta = \tilde{P} \times \Psi$, $[\mu,\nu,\xi] = \vartheta / (\vartheta \cdot \vartheta)^\half$ representing a `unit vector north'.
Vectors in the local patch coordinate system can be transformed into vectors in the global coordinate system by noting that each component in the global coordinate system is the scalar product of the local vector with the global unit vectors, which encourages that the local vectors are considered to be column vectors, while global vectors are considered to be row vectors. The transformations may thus be written
\begin{equation}
    \begin{bmatrix}
    g_1 & g_2 & g_3 
    \end{bmatrix}
    = \rot \cdot
    \begin{bmatrix}
    l_1 \\
    l_2 \\
    l_3
    \end{bmatrix} \,,
\end{equation}
while the transformation from global to local coordinates correspondingly is
\begin{equation}
    \begin{bmatrix}
    l_1 \\
    l_2 \\
    l_3
    \end{bmatrix}
    = 
    \begin{bmatrix}
    g_1 & g_2 & g_3 
    \end{bmatrix} 
    \cdot \rot \ .
\end{equation}
In Fortran and Python these operations can be performed using the built-in \texttt{matmul(a,b)} procedures, which allow both \texttt{a} and \texttt{b} to be either vectors or matrices.

If coordinates in the global and local coordinate systems share a common origin these transformations are complete. If they are relative to the centres of the local patches, the distance between the patch centres need to be taken into account, by being added either before (if given in the global coordinate system) or after the matrix multiplication (if given in the local coordinate system).

\subsubsection{Relative coordinate transformations}
To transform from one local coordinate system to another, for example in the context of guard zone interpolations, one could transform from one local system to a global system and then from there to another local system.  It is more convenient, however, to make use of a \textit{relative transformation matrix}, which has the local unit vectors of the `source' patch, expressed in the `target' coordinate system as columns, and correspondingly has the local unit vectors of the `target' patch expressed in the `source' coordinate system as rows.  It is clear then that \textit{the `source' coordinates take the role of the `local' coordinates above, while the `target' coordinates corresponds to the `global' coordinates above} (common to a number of different `sources' that contribute to the same `target'). 
\begin{equation}
    \label{eq:src_to_tgt}
    \begin{bmatrix}
    t_1 & t_2 & t_3 
    \end{bmatrix}
    = \rot \cdot
    \begin{bmatrix}
    s_1 \\
    s_2 \\
    s_3
    \end{bmatrix} \,,
\end{equation}
and
\begin{equation}
    \label{eq:tgt_to_src}
    \begin{bmatrix}
    s_1 \\
    s_2 \\
    s_3
    \end{bmatrix}
    = 
    \begin{bmatrix}
    t_1 & t_2 & t_3 
    \end{bmatrix} 
    \cdot \rot \ .
\end{equation}
Since the transformation from source to target can also be achieved by first translating the source coordinates to global coordinates and then go from there to target coordinates, it follows that the relative transformation can be obtained as the product of the transposed target rotation matrix and the source rotation matrix:
\begin{equation}
\label{eq:rot_combi}
    \rot = \rot_\tgt^T \cdot \rot_\src
\end{equation}

\subsection{Guard zone interpolations}

\begin{figure}
\centering
    \includegraphics[width=0.85\columnwidth]{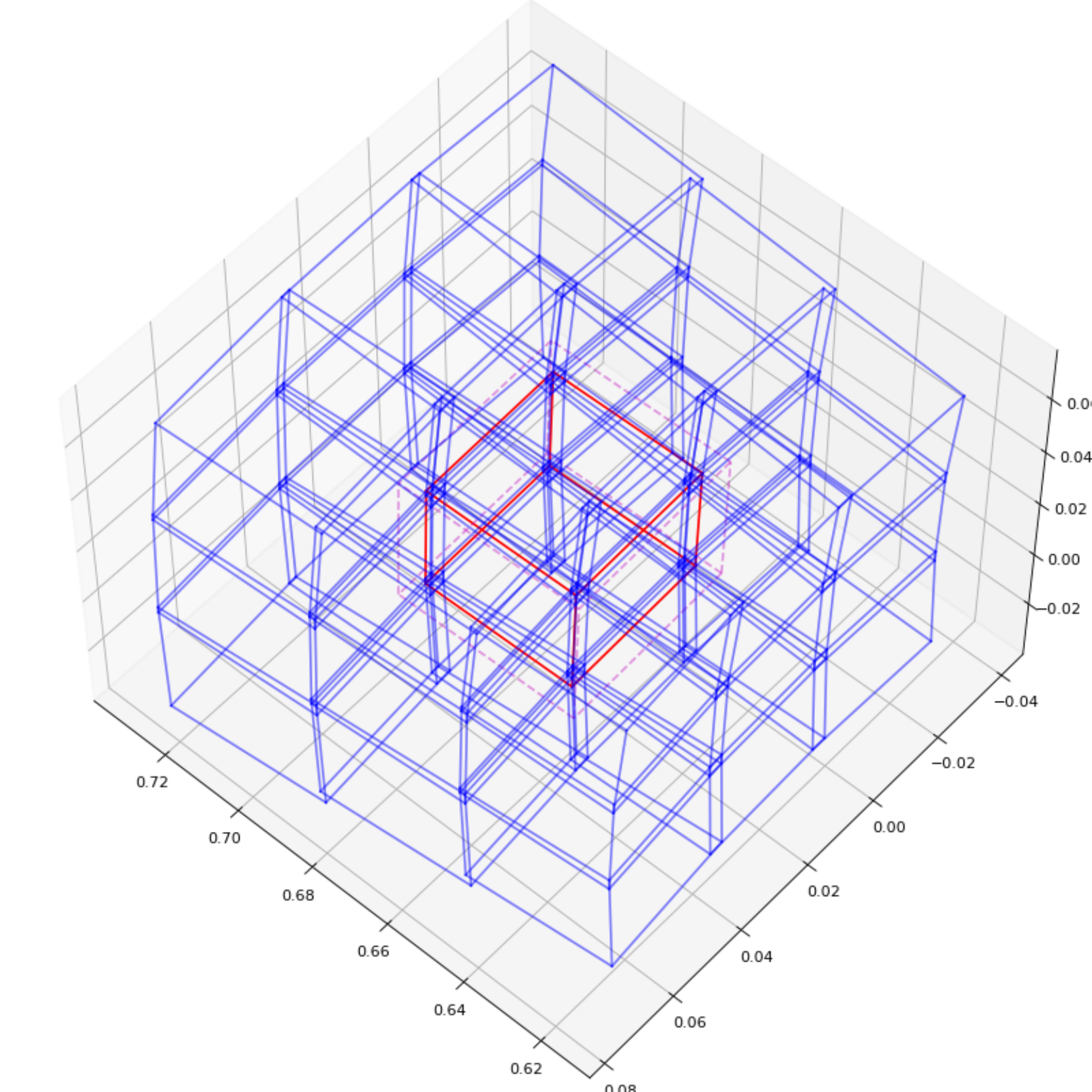}
    \caption{A \texttt{target} patch (red) and its \texttt{source} neighbours (blue). Red dashed lines show the guard zones of the \texttt{target}. Note the slight tilt between adjacent patches.}
    \label{fig:vb_patches}
\end{figure}

DISPATCH utilises local time-stepping for individual patches, thus each patch needs guard zone data from neighbour (\texttt{nbor}) patches interpolated in both space and time. Since the mass density varies a lot (both vertically and horizontally), we convert to per-unit-mass variables (log density, velocity, and entropy per unit mass), where applicable, before doing time-interpolations into the \texttt{target's} time. Most patches have only slightly tilted neighbours, making it potentially possible to use standard Cartesian coordinate interpolations, where only the fractional interpolation weight varies from point to point. 
However, near the `seams' between the six volleyball faces, the patches can have arbitrary overlaps and different north vectors, which require more general interpolation methods. In this context we use the concepts of \textit{regions of authority} (ROA) and \textit{regions of interest} (ROI). The region of authority of a patch is the volume inside which the patch ``has authority'' and updates the variable values.  The region of interest in addition includes the guard zones of the patch, and potentially may include sub-volumes where a neighbour overlapping patch has a higher level of refinement.
For simplicity and consistency we currently use these more general point-by-point interpolation methods for all patches.
\Figure{vb_patches} shows a \texttt{target} patch with its surrounding \texttt{source} neighbours. Note the slight tilt between patches.  

For patches with a significant relative tilt between each other, a simple lower-left-corner and upper-right-corner interval range for the ROI is not sufficient. Instead, we take all 8 corner points and find the smallest and largest index in the local frame of reference of the patch we need, to get guard zone data for the \texttt{target}.

\subsubsection{Interpolation in time}
Each patch stores the MHD quantities at 5 different times, in a rotating time-slot buffer. As described in \cite{Nordlund_2018MNRAS}, a patch is ready to be updated when all \texttt{nbor} patches are ahead in time (including a potential `grace' time interval) of \texttt{target}. We perform linear interpolation between the time slots of \texttt{nbor} (\texttt{source}) to find the time slot indices and weights. These weights are then applied to the selected \texttt{source} memory slots (using only the relevant index sub-regions) to get the required quantities for the given \texttt{target} time. Higher order interpolation in time is available as an option.

\subsubsection{Cell-centred quantities}
The rotation matrix $\rot$ is defined at the centre of the patch, thus to get the \texttt{target} cell coordinates $p_t=[x_t,y_t,z_t]$ in the \texttt{source} system $p_s=[x_s,y_s,z_s]$, we need to take the scalar product of the location in the \texttt{target} system with the rotation matrix $\rot$ of the \texttt{source} expressed in \texttt{target} coordinates. These are available from \Eq{rot_combi}. The coordinates that go into the operation are the \texttt{target} coordinates relative to the \texttt{source} position, i.e.\ the distance from \texttt{target} to \texttt{source}, $d$ is subtracted before the multiplication:
\begin{equation}
    \label{eq:interp_indices}
    p_s = [(p_t - d) \cdot \rot]/c + o,
\end{equation}
where $c$ is the patch's cell size and $o$ is the patch's position offset in index space.
We go through the \texttt{target's} ROI and map all the \texttt{target} points to \texttt{source} points in its respective ROA. Of course, not all ROI might be available in a given index space we previously determined, thus we use an array, which tells us if a given data point is \texttt{available}.

Lastly, we re-interpolate the MHD quantities from the \texttt{source} to the \texttt{target's} frame of reference. For scalars this is just a simple interpolation in space, but for vectors (e.g.\ cell-centred velocities) we need to transform the vector components to the \texttt{target} system by using \Eq{src_to_tgt}.

\subsubsection{Face-centred quantities}
Face-centred vector fields, e.g.\ the magnetic field, pose an additional complication---different vector components are located on different faces. Equation \ref{eq:interp_indices} now needs an additional component for the staggered position of the vector:
\begin{equation}
    \label{eq:interp_indices_stag}
    p_s = [(p_t - d) \cdot \rot]/c + o + h_\mu,
\end{equation}
where $h_\mu$ is the staggering for the given component with $\mu$ indicating a coordinate axis. This gives us three position data sets for the vector components. We go through the \texttt{target's} ROI and map all the \texttt{target} points to \texttt{source} points in its respective ROA. For each vector component, we re-stagger the perpendicular vector components into the position of the vector component in question and store them separately. Later these re-staggered components are used to transform the vector to the \texttt{target} system of reference by using \Eq{src_to_tgt}.


\section{Pilot experiment design} \label{sec:experiment_design}
Table \ref{tab:sim_setup} summarises the basic parameters for the mesh decomposition we use. The  simulation spans 0.655--0.998 R$_\odot$. With N$_\theta$ = 96, we have 19 layers in the setup. The maximum resolution near the surface is 194km. This is sufficient to resolve the small scale flows at this depth and to relax the initial setup into a steady-state convective Sun in the shortest time possible. 

\begin{figure}
\centering
    \includegraphics[width=0.85\columnwidth]{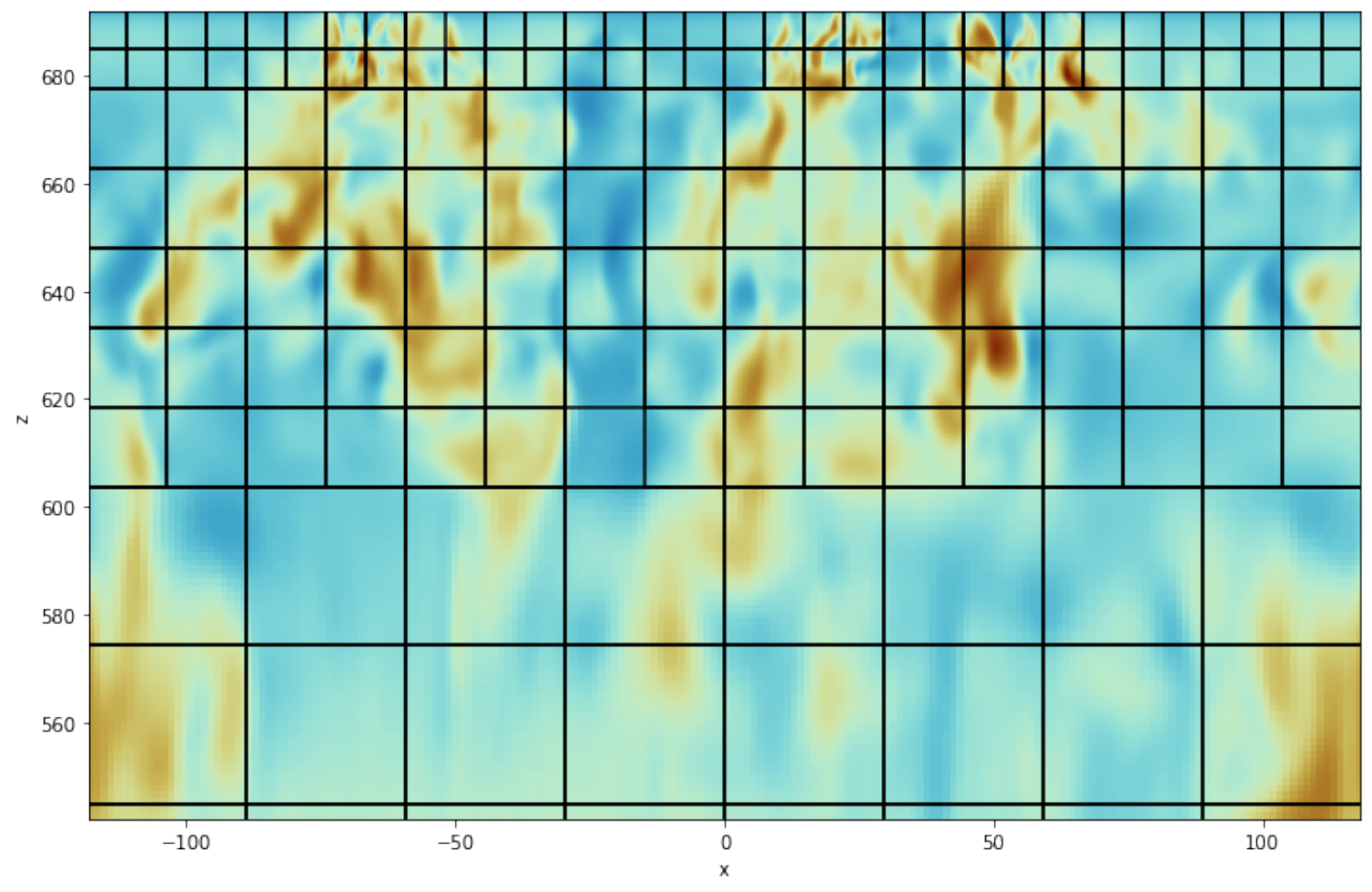}
    \caption{Static mesh refinement in a 'sandbox' version of the simulation. 2 additional levels are added, starting at 0.88 and 0.97 R$_\odot$.}
    \label{fig:mesh_refinement}
\end{figure}

In future simulations we will be able to further increase the resolution by using static or adaptive mesh refinement to focus on particular features of interest, e.g. magnetic flux emergence, convective down-flows, etc. An example of static refinement is shown in Figure \ref{fig:mesh_refinement}. Here, only 2 levels of refinement are added at 0.88 and 0.97 R$_\odot$. Normally, we do a factor of 2 refinement, creating 8 new \texttt{leaf} patches on top of a \texttt{parent} patch. Newly created patches are co-aligned with the parent patch, so their $\rot$ is the same as \texttt{parent's}. 

The top layers are the most numerous and each additional level of refinement means a factor of 2 reduction in time-step and 8 times more cells, so we don't want to run with maximal resolution the whole time. Such simple mesh refinement can be added during routine restarts of the run. 

In Appendix \ref{sec:photosphere} we demonstrate a further refinement of the experiment down to 71 km cells near the surface when targeting the photosphere. The refinement beyond that scale is only sensible using adaptive mesh refinement (AMR), targeted on features of interest.

\begin{table}
\caption{Configuration of the Pilot experiment.}            
\label{tab:sim_setup}      
\centering                         
\begin{tabular}{l c}        
\hline                       
   Bottom radius & 0.655 R$_\odot$ \\      
   Top radius & 0.998 R$_\odot$ \\
   N$_\theta$ & 96 \\
   N$_{\theta, min}$ & 24 \\
   N$_{\theta}$ reduction rate & 40, 8, 8, 4, 4, 4, 4 \\
   Corresponding levels of refinement & 3\\
   Smallest cell size near surface & 194 km \\
   Largest cell size at 0.898 R$_\odot$ & 705 km \\
   Time unit & $10^3$ s \\
   Length unit & $10^6$ m (1 Mm) \\
   Mass unit & $10^{23}$ g \\
   $\Omega_{coriolis}$ & 424 nHz \\ 
\hline                                  
\end{tabular}
\end{table}

In parallel with the primary Pilot simulation, we ran two additional classes of simulations, with increased simplifications. The first one --- the `small volleyball' has exactly the same parameters and the same physics, but the resolution is lower, with 1.56 Mm cell size at the finest level. It contains about 30,000 patches.

The second class, denoted `sandbox' runs, is even more simplified:
\begin{itemize}
    \item the volleyball geometry is replaced with a plane-parallel Cartesian mesh decomposition;
    \item the vertical and horizontal extent depends on the goals of the particular `sandbox' run, with typically the vertical extent the same as the main experiment, and a horizontal extent of order 100-300 Mm with periodic boundaries;
    \item a Coriolis force is not present;
    \item all other physical effects are identical;
    \item AMR can be used to reach any desired resolution.
\end{itemize}
These two simulations accompany the main one as test-beds. They run much faster and are much cheaper. The `sandbox' simulation can run on a local workstation, with just a handful of cores assigned to it, whereas comparable resolution simulations, e.g. \cite{Hotta_2019,Hotta_2020} requires an HPC allocation. Local time-stepping removes the global CFL constraint set by the deepest, hottest cells. With three levels of refinement the smallest cell size in this simulation is 123 km, which is better than the 198 km horizontal resolution achieved in \cite{Hotta_2019}. Adding a photosphere on top requires one or two additional levels of refinement with smallest cell size being 61 km or 30 km respectively. This last setup then requires a few compute nodes to be executed efficiently. Similarly, the `small' simulation can be run on a single compute node. The main role of these accompanying simulations is to run far ahead of the main simulation, and to validate the physics modules, the stability of the setup, etc.

Numerical simulations of the Sun can be as complicated as one wants and can provide a wealth of different diagnostics (e.g.\ \citealt{Przybylski_2022,Carlsson_2016,Kohutova_2021,Finley_2022,Druett_2022} to name a few), but for now, we limit ourselves to only a few key variables. Additional micro- and macro-physics effects can be added later.

\subsection{Gravity and rotation}
Gravity is radial, evaluated in global coordinates and transformed to each patch frame for each individual cell for the local update: 
\begin{equation}
 \mathbf{g}(r) = -g(r)\mathbf{\hat{r}},   
\end{equation}
where $\mathbf{\hat{r}} = [x,y,z + r_{\rm{c}}]$, with $[x,y,z]$ being the cell positions in the local frame of reference, relative to the centre of the patch, $r_c$ is the radial distance from the centre of the Sun to the centre of the patch, and $r=\sqrt{x^2+y^2+(z+r_c)^2}$ is the radial distance from the centre of the Sun to a given cell. 

If rotation is included, the rotation vector $\mathbf{\Omega}_{\rm{coriolis}}$ is constant in the inertial frame and the corresponding inertial accelerations (from Coriolis and centripetal forces) are included as source terms in the momentum equation (Section \ref{sec:numerical_scheme}). The chosen rotation rate used in the pilot is listed in Table \ref{tab:sim_setup}.

Alternatively, one can initialise the experiment with a differential rotation already present (e.g. with iniital angular velocity taken from HMI, \citealt{Larson_2018}) and then track the deviations from the initial profile.

\subsection{Thermodynamics and initial hydrostatic reference state} \label{sec:initial_hse}
The thermodynamics are provided by the EOS described in Section \ref{sec:EOS}. We construct an initial spherically symmetric reference stratification that is close to hydrostatic equilibrium under the adopted EOS and gravity. In practice, we start from a standard solar 1D model (e.g. Model-S from \cite{JCD_1996}) for $(\rho,T)$ as a function of radius. We re-integrate the hydrostatic balance to be fully consistent with our adopted EOS with a given element abundance. Additionally, we shift the entropy by a constant value to have the flat, adiabatic part of the convection zone at 0 - this gives the best numerical precision.

To accelerate symmetry breaking and the onset of convective flows, we perturb only the velocities on globally coherent scales, normalised to patch size:
\begin{equation}
    \delta u = f_p e^{-(dr/w_p)^2},
\end{equation}
where $f_p \in [-1,1]$ is a random number, $dr$ is a distance from a centre of a patch, and $w_p$ is perturbation diminishing weight.  The resulting noise $\delta u$ is then smoothed by averaging over 7 cells in the two horizontal directions and corrected for curvature:
\begin{equation}
\mathbf{u_{xyz}} = \mathbf{u_{xyz}}+\sum_{x,y=-3}^3 \left[ \delta u_{x,y} \hat{r} \right].
\end{equation}
Lastly, we normalise the perturbation to have the same amplitude everywhere.

\subsection{Surface cooling}
\label{subsubsec:surface_cooling}
Solar convection is driven by cooling in the top layers. In this Pilot experiment setup we use parametrized surface cooling, as the upper boundary is below the photosphere, while explicit, ray-based radiative heat transfer is preferred in simulations where the photosphere is included.

We emphasize that the parametrized surface cooling described here is not an alternative to or replacement for ray-based radiative transfer.  When the surface is included ray-based radiative transfer should always be used; it does increase the cost per cell in the layers where it is being used, but this is always a small factor relative to the total update cost, so it would be pointless to use a parametrized cooling in such a context.  On the other hand, when the radiating surface is not included, it is simply not possible to use ray-based radiative transfer, and hence the surface cooling must be represented by a parametrized model, such as the one being described here.

In each patch adjacent to the upper boundary, the entropy per unit mass is relaxed towards a target value
\begin{equation}
  S_\mathrm{target} = S_\mathrm{1D}(r_\mathrm{top})
                      - \psi\,\left(S_\mathrm{max} - S_\mathrm{floor}\right),
\end{equation}
using
\begin{equation} \label{eq:surf_cool}
  S^{n+1} = S^{n} + \left(S_\mathrm{target} - S^{n}\right)
            \frac{\Delta t}{\tau_\mathrm{top}}\,e^{-d/h_b},
  \quad d = \max\left(r_\mathrm{top} - r,\,0\right),
\end{equation}
where $S_\mathrm{1D}(r_\mathrm{top})$ is the entropy of the initial stratification at the top interior radius $r_\mathrm{top}$, $S_\mathrm{max}$ is the maximum entropy in the bottom interior layer of the patch, $S_\mathrm{floor}$ is the entropy of the coldest descending material, calibration of which is shown in Appendix \ref{sec:surf_cool_calib}, $\Delta t$ is the local time step, $\tau_\mathrm{top}$ the cooling timescale, and $h_b$ the e-folding depth of the cooling below $r_\mathrm{top}$. 

Equation~ \ref{eq:surf_cool} combines relaxation of each cell towards $S_\mathrm{1D}(r_\mathrm{top})$ on the timescale $\tau_\mathrm{top}$ with an entropy sink $\psi(S_\mathrm{max}-S_\mathrm{floor})/\tau_\mathrm{top}$. The dimensionless parameter $\psi$ sets the cooling strength in units of the entropy contrast available in the patch; we use $\psi = 1$ here. In the Pilot setup, $S_\mathrm{max}$ is evaluated $\approx 4.7$\,Mm below $r_\mathrm{top}$, where
it samples the near-adiabatic upflows. The cooling acts on upflows and downflows alike. No lower bound
is imposed on $S$, and the emergent luminosity is not explicitly prescribed, but the calibration, described in Appendix \ref{sec:surf_cool_calib} is meant to match the solar luminosity.

Models that include the optical surface incorporate full radiative transfer with multi-bin opacities, which refine the cooling dynamics further and allow for a more realistic representation of energy transport. We use the conservative formulation for radiative transfer \citep{Eitner_2025}.

\subsection{Initial magnetic field} \label{subsec:init_b_field}

There are many ways to initialise the magnetic field in local, Cartesian simulation domains, e.g.\ uniform vertical, salt-and-pepper, flux emergence, etc. There are fewer ways to initialise a magnetic field in a global solar simulation. A main requirement is that it must be divergence free; the Sun must not become a magnetic monopole (i.e.\ no purely radial magnetic field). Thus the magnetic field can be initialised either as a torus, or as a uniform field in the global frame of reference. We initialise a uniform magnetic field in a 'hollow cylinder' shape, aligned with the vertical axis in the global frame of reference, i.e.\ $\bf{B}_0=[\rm{B_x}, B_y, B_z] = [0, 0, 0.1]$ Gauss. The initialised constant magnetic field is then transformed to local frame of reference via \Eq{src_to_tgt}.
The magnetic field is truncated at a particular $r_{min}$, which we take as the bottom of the simulation box. The reasoning is simple - we don't want vertical magnetic field permanently frozen near the centre of the simulation domain -- there are no motions to advect the field. The cylindrical void, aligned with the rotation axis, is eventually filled in by the advected magnetic fields from the other parts of the Sun. 

We chose such an approach for two reasons:
\begin{enumerate}
    \item simplicity--$\nabla \cdot B=0$ by definition initially;
    \item convective motions quickly erase the simplicity of this primordial magnetic field alignment.
\end{enumerate}


\begin{figure*}
  \centering
  \resizebox{\hsize}{!}
  {\includegraphics[width=0.99\columnwidth]{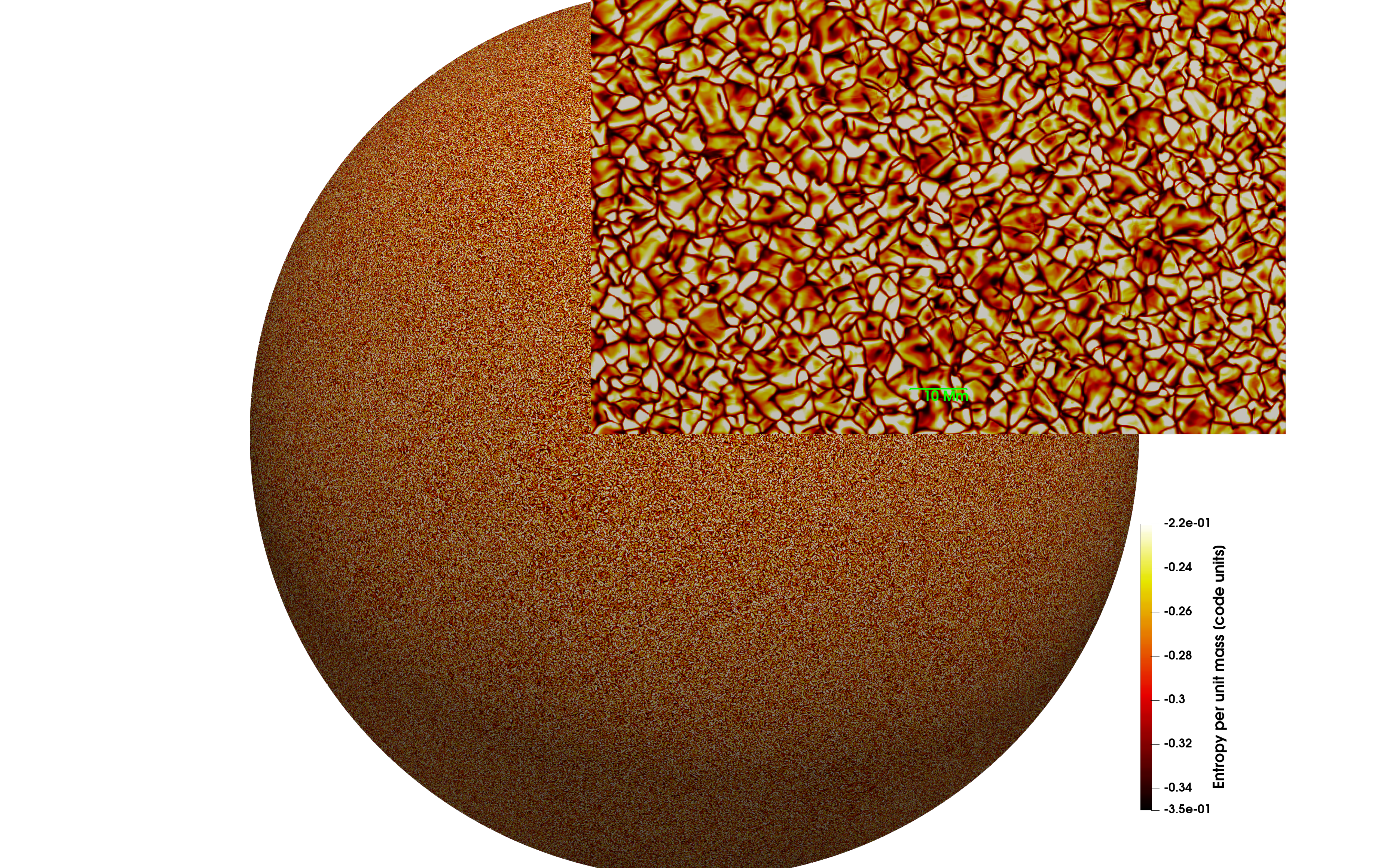}}
  \caption{Entropy per unit mass in code units 500 km below the top boundary in the simulation. A zoom-in with a 10Mm scale is shown at the top right to make the convection patterns visible. Movie available online.}
  \label{fig:vb_S_surface}
\end{figure*}

\subsection{Boundary conditions}
\label{subsec:bcs}

The experiment covers the entire surface of the Sun, so only radial boundary
conditions (BCs) need to be considered.

Both radial boundaries are constructed to be \emph{stateless}, i.e. the complete guard-zone state is rebuilt from the last interior layers on every update. No boundary quantity depends on the guard-zone contents left by a previous update, on the restart history, or on the states computed by the Riemann solver. In a task-based code with asynchronous, per-patch time stepping this guarantees that a given interior state maps to a unique guard-zone state irrespective of the order in which patches are updated.

\subsubsection{Bottom boundary}
\label{subsubsec:bottom_bc}

In the Pilot setup the bottom boundary is located at $r = 0.655\,R_\odot$, below the tachocline, where there are no convective motions, while some waves can still propagate.

\paragraph{Momenta.}
All three momentum components are reflected symmetrically about the last internal layer. For the horizontal components this is a stress-free (no-shear) condition. For the radial component the symmetric reflection leaves the boundary open, so that mass may cross it in either direction, e.g. the downflows reaching the bottom of the domain must be allowed to leave rather than be reflected back into the convection zone.

\paragraph{Thermodynamic variables.}
At the last internal layer, inflows and outflows are distinguished by the sign of the radial momentum there. Where material enters the domain, the density and the entropy per unit mass are relaxed towards their initial values $\rho_0$ and $S_0$ with a characteristic time $\tau_{\rm bot} = 1000$~s; where material leaves, the interior values are retained.

In the guard zones the two cases are treated separately. Incoming material is assigned the initial profile exactly, $\rho = \rho_0$ and $s = S_0$, so that the specific entropy of material entering the domain is fixed. For outflow columns the guard zones are given the initial profile plus the current deviation at the last internal layer,
\begin{equation}
\label{eq:bottom_outflow}
    S(z) = S_0(z) + \left[\, S(z_{\rm i}) - S_0(z_{\rm i}) \,\right],
\end{equation}
where $z_{\rm i}$ denotes the last internal layer. In both cases, the guard-zone density is set to $\rho_0$, which anchors the guard-zone pressure to the initial stratification and thereby provides a weak restoring force
against secular drift of the total mass in the domain.

In principle, the specific entropy at the bottom boundary, together with the surface cooling ultimately determines the entropy contrast that drives the convection, and hence the energy flux carried through the convection zone.  In practice, the Kelvin-Helmholtz time of the convection zone is so large that the entropy at the bottom must be considered to be given by the initial conditions, which are taken from a standard solar model, e.g. Model-S \citep{JCD_1996}.

\paragraph{Magnetic field.}
The tangential field components are damped towards zero over a length $\ell_{\rm damp}$ of a few grid cells, anchored at the last internal layer. The horizontal mean is damped along with the fluctuations: the reservoir below the boundary is assumed to be field free, so that the boundary is strictly lossy for tangential flux. Outflows advect field out of the domain at full interior strength, while inflows can only carry back the exponentially reduced remnant of what the interior itself placed at the boundary interface, and no prescribed field pattern exists anywhere to be fed in. A tachocline-like imposed field can be introduced in future experiments by damping towards a prescribed profile rather than towards zero.

The radial component is not damped, since damping the normal component would violate flux conservation through the boundary and hence $\nabla\cdot\bm{B}=0$. Instead, it is reconstructed face by face by downward recursion of the staggered solenoidal constraint: each guard-zone face value is chosen so that the divergence of the cell above it vanishes exactly, starting from the outermost internal face. The construction is exact to machine precision, so no iterative cleaning is required. The bottom boundary is a complete spherical shell with no lateral edges (or is horizontally periodic in `sandbox` simulations), so the net radial flux through it vanishes identically for any solenoidal field and is preserved to machine precision by the constrained-transport evolution independently of this fill, thus the radial component requires no prescription of its own.

\subsubsection{Top boundary}
\label{subsubsec:top_bc}

The top boundary is cork-like: the guard zones are filled with a stratification that is close to hydrostatic and dampens the velocity field, so that the boundary is nearly closed while still permitting the domain to relax. In the Pilot experiment, it is located at $\approx$1.4 Mm below the surface, but the boundary conditions are constructed to also accommodate the simulations that extend beyond the photosphere. 

\paragraph{Density.}
The guard-zone density is obtained by extrapolating hydrostatically from the last internal layer, with the adiabatic scale height $\gamma H_P$ frozen at its value there,
\begin{equation}
\label{eq:top_density}
    \frac{\rho(z)}{\rho(z_{\rm i})} =
    \exp\!\left( \frac{(z - z_{\rm i})\,\rho(z_{\rm i})\,g}
                      {\gamma P_{\rm gas}(z_{\rm i})} \right) \, f ,
\end{equation}
modulated by a flow-dependent factor
\begin{equation}
    f = {\rm min} \left({\rm max} \left(1 + u_z/c_{\rm s}, f_{\rm min}\right), f_{\rm max}\right),
\end{equation}
evaluated at the last internal layer. The factor is the zero-incoming-wave condition $\delta\rho/\rho = u_z/c_{\rm s}$ and thus the boundary is weakly transmitting. The clipping bounds the response for strong shocks and for strong drainage. The result is capped at the value at the last internal layer, so that a density inversion, against which material could accumulate, should never form in the guard zones.

\paragraph{Entropy.}
In the experiments with upper BC below the optical surface, the entropy per unit mass is held constant through the guard zones, implying adiabatic vertical temperature gradients. The surface cooling described in Section \ref{subsubsec:surface_cooling} is then applied to the entropy per unit mass.

In patches that reach the photosphere, where the entropy per unit mass drops sharply and then begins to rise again with height (e.g.\ dut to entropy production from magnetic field braiding and shocks), this is supplemented by a monotonicity-limited extrapolation of the entropy  in outflow columns. The reach of that extrapolation is limited to $n_{\rm slope}=[0,1]$ cells irrespective of the guard-zone depth. With $n_{\rm slope}=1$ it is blended smoothly with the constant-entropy state over a narrow band of vertical velocities around zero. The extrapolation closes a feedback loop between the guard-zone entropy, the pressure of the overlying layers, and the vertical velocity at the boundary, and without a bound on its gain, the loop may become unstable. Setting $n_{\rm slope}=0$ recovers the plain constant-entropy condition, which is the zero-gain limit.

\paragraph{Momenta.}
The horizontal components of the momenta are tapered with $\rho^{3/2}$, consistent with a no-shear boundary condition on the horizontal velocities and preventing unphysical growth of the kinetic energy density in the tenuous guard-zone cells. The radial momentum is tapered with the square of the density ratio. The radial kinetic energy density in the guard zones therefore falls as $\rho^3$ and horizontal as $\rho^2$ outwards, so that the guard zones in both cases are energetically inert and cannot return energy to the domain. This is the principal sink for wave reflection at a non-characteristic boundary.

\paragraph{Magnetic field.}
The tangential components are damped over a length $\ell_{\rm damp}$ of a few grid cells, but --- in contrast to the bottom boundary --- the horizontal mean is preserved and only the fluctuating part decays. This reproduces the behaviour of a potential-field continuation, in which each horizontal mode decays as $\exp(-k\,\Delta z)$ and the $k=0$ mode does not decay at all; damping the mean instead would place a current sheet at the top of the domain whenever the field carries net horizontal flux. The radial component is reconstructed by upward recursion of the staggered solenoidal constraint, in the same manner as at the bottom boundary.

In future simulations, when the domain is extended into the chromosphere and beyond, potential-field extrapolation will be required for the upper magnetic boundary condition, and the characteristic decomposition of the hydrodynamic variables will become necessary as the magnetic field becomes dynamically dominant at the top of the domain.

\begin{figure*}
  \centering
  \resizebox{\hsize}{!}
  {\includegraphics[width=0.99\columnwidth]{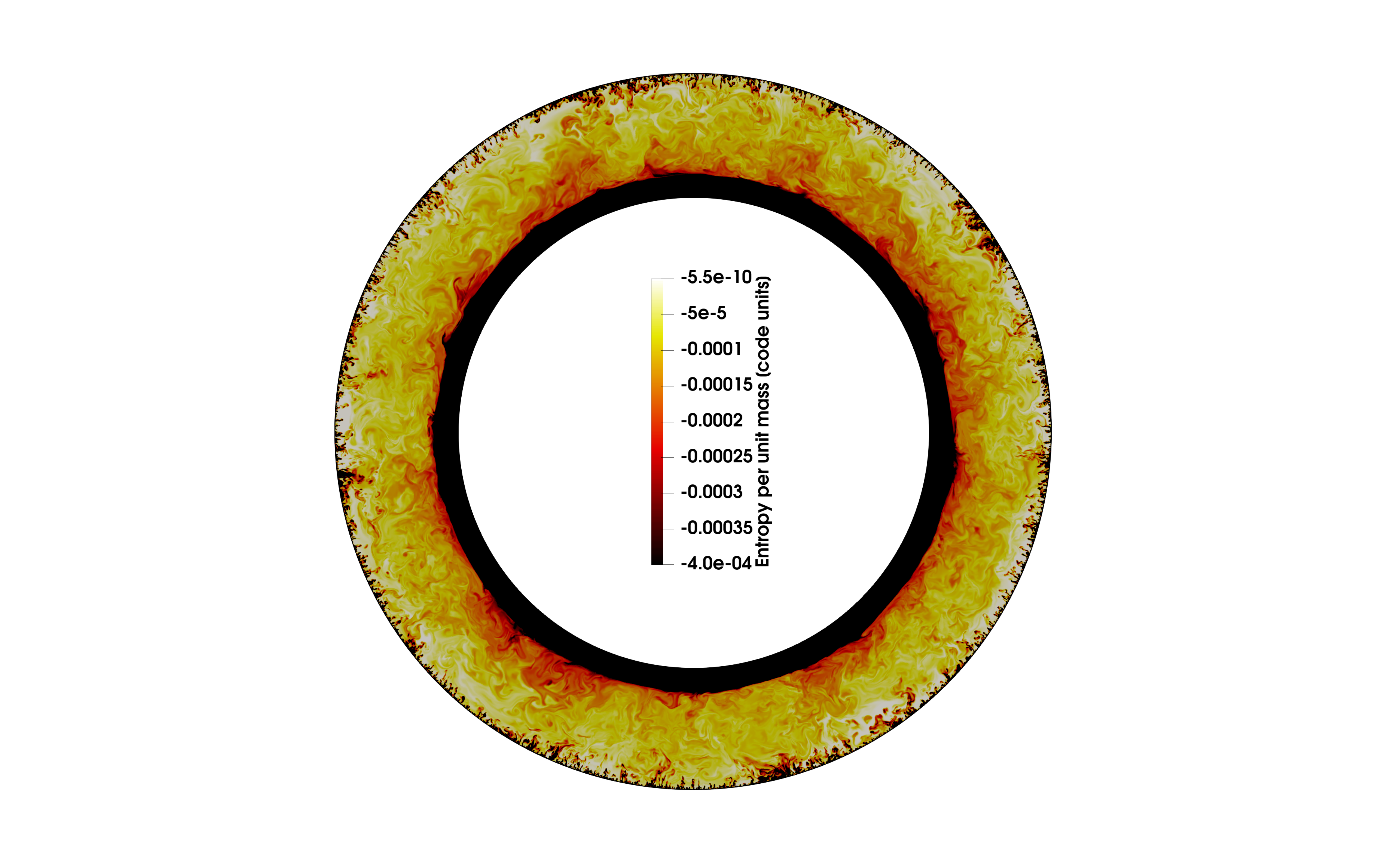}}
  \caption{Slice across the simulation domain at the prime meridian, showing entropy per unit mass.}
  \label{fig:vb_S_slice}
\end{figure*}

\section{Relaxation and initial results}
\label{sec:relax}
Due to finite numerical precision we expect the initial hydrostatic equilibrium not to be perfect and some oscillations might occur while the system is relaxing into an quasi-equilibrium state. For the first 5 time units (5000 solar seconds) we apply Rayleigh drag on all momentum components. Additionally, for the first 20 time units (20 000 solar seconds) we use wave-killing to damp the waves near the upper boundary. We purposively do not convert the lost kinetic energy due to drag and wave killing into thermal energy. After the said 20 time units we allow the system to evolve freely without any imposed damping.

We see the first signs of convective motions appearing at around 2 hours of solar time. However, it takes a lot longer for the low entropy down-flows to penetrate all the way from the surface to the deep convection zone.

Figure \ref{fig:vb_S_surface} shows entropy per unit mass 500 km below the top boundary (around 2 Mm below the solar surface). The insert is a zoom in to show the actual granules. There, a 10 Mm scale is given as well. We can see, that the typical size of the convective cells is of the order of 2-4 Mm, with some being a bit larger. Figure \ref{fig:vb_S_slice} shows a slice through the domain at the prime meridian. ``Entropy rain'' penetrates deeply into the convection zone. The base of the convection zone can be identified from the sharp drop in entropy per unit mass, convection is fully developed. No shear layer has formed yet.

\begin{figure*}
  \centering
  \resizebox{\hsize}{!}
  {\includegraphics[width=0.99\columnwidth]{./figures/pdf/vr_sphere.pdf}}
  \caption{Radial velocity in km/s 200 km below the top boundary in the simulation. A cone-slice with a 10 Mm scale is shown as well.}
  \label{fig:vb_uz}
\end{figure*}

Figure \ref{fig:vb_uz} shows radial velocity 200 km below the top boundary in the simulation. Typical radial velocities for up-flows are of the order of a 100 - 300 m/s, and 1.5 km/s for down-flows, with the extrema of ~1.5 km/s for up-flows and 3 km/s for down-flows.  The cone cut-out shows that the typical convection scales become increasingly large with smaller radius, as expected. The velocity vector fields do not show any interpolation glitches between adjacent patches. This also includes seams, where relative angles are large and we need to count on accurate transformations. We took a random sample of individual patches throughout the box and investigated the transition between the guard zones and interiors just after the guard zone interpolation. The transitions in both vector and scalar quantities were smooth, without any discontinuities. This is very encouraging, as smooth transitions between interfaces can be rather challenging (e.g.\ \citealt{Brchnelova_2022}).

\begin{figure*}
  \centering
  \subfigure[Surface 4.5 Mm below the top boundary.]{\includegraphics[width=0.99\columnwidth]{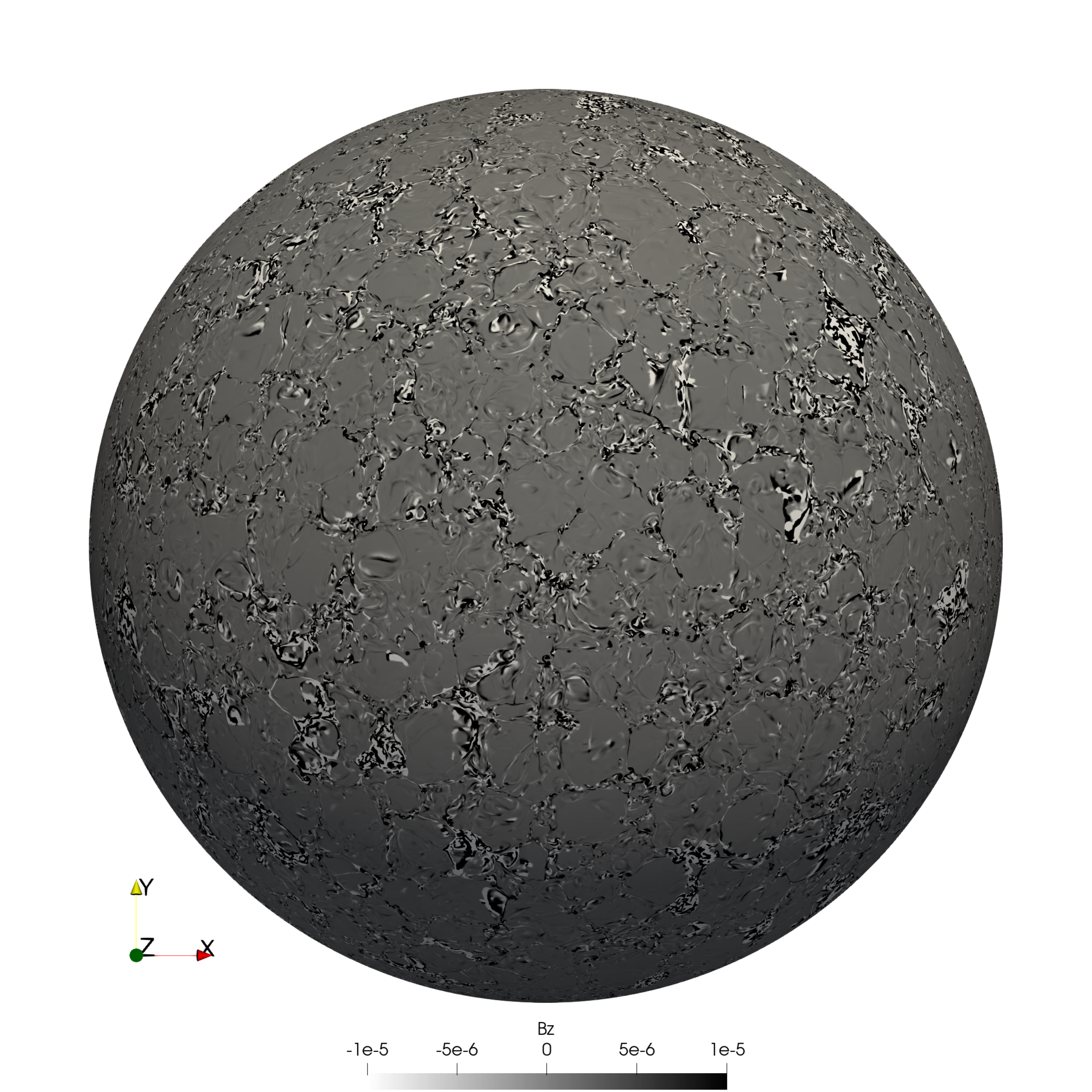}}
  \subfigure[Slice across the simulation domain at the prime meridian.]{\includegraphics[width=0.99\columnwidth]{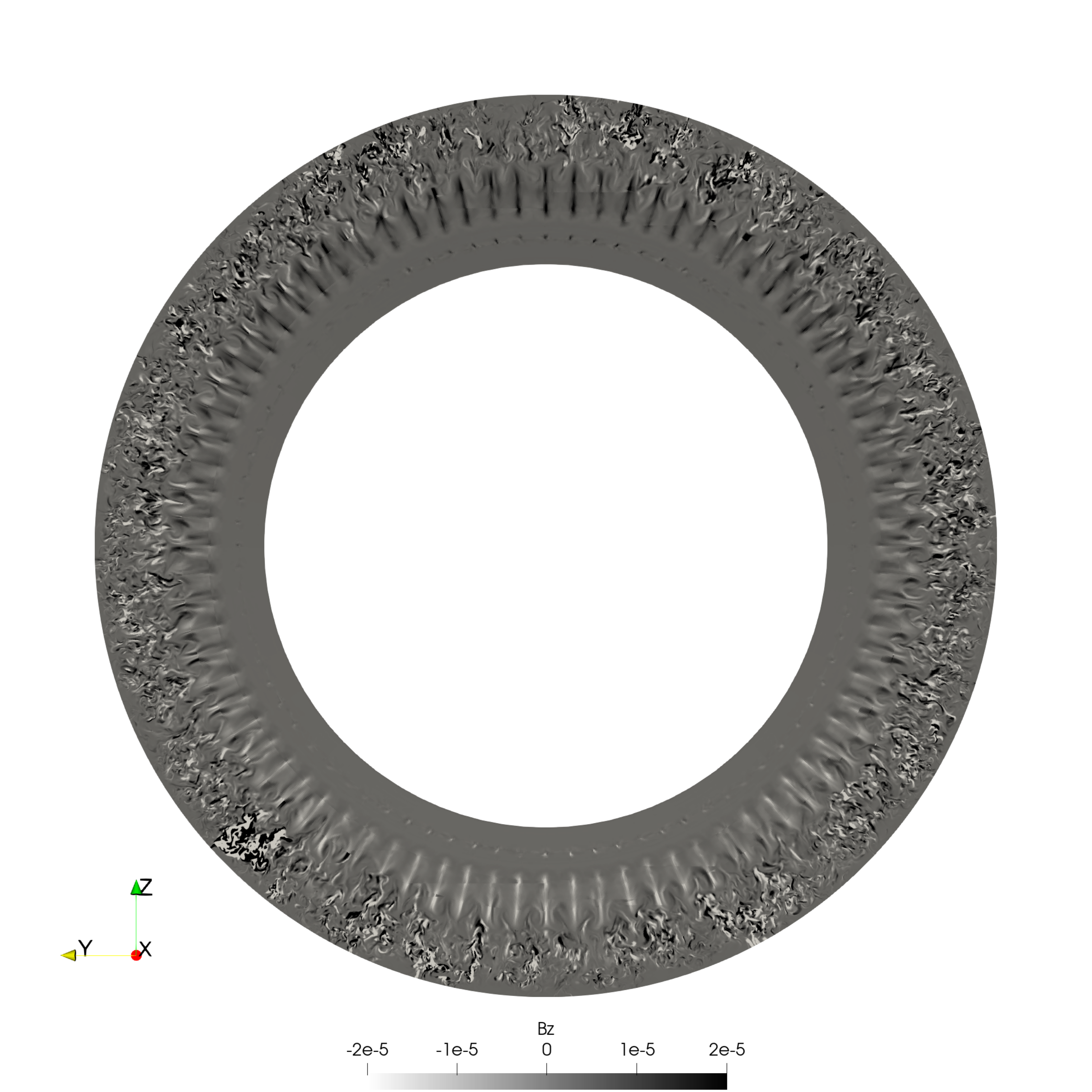}}
  \caption{Radial magnetic field component, in code units.}
  \label{fig:vb_bz}
\end{figure*}

Using our HLLS solver provides us with a unique bug-catching tool. Guard zone interpolation glitches create artificial discontinuities. Discontinuities trigger jump conditions, which in turn increase entropy per unit mass. Even errors in the magnetic field can be misidentified as magnetosonic waves, which can increase entropy. Even if the build-up of entropy per unit mass is marginal, after tens of thousands of updates and continuous build-up we would be able to see glitches in entropy per unit mass, even if other quantities might be able to partially ``mask'' them. Neither of the figures \ref{fig:vb_S_surface} and \ref{fig:vb_S_slice} show any signs of such localised entropy increases.

\Figure{vb_bz} shows the radial component of the magnetic field in code units. From both the left and right hand side panels it can be seen that the primordial structure is already mostly hidden by local dynamo effects; in some places the magnetic field magnitude is already over two orders of magnitude stronger than the initial mean magnetic field. We can, in fact, see some magnetic vortices forming in some areas. These initial concentrations could later give rise to magnetic activity at the surface, e.g.\ in the forms of network structure, pores, and sunspts. The local dynamo effects are still weak due to short time of evolution, but the topological and morphological effects of the convective flow hierarchy are expected to be largely independent of magnitudes.

\begin{figure*}
  \centering
  \subfigure[Density]{\includegraphics[width=0.66\columnwidth]{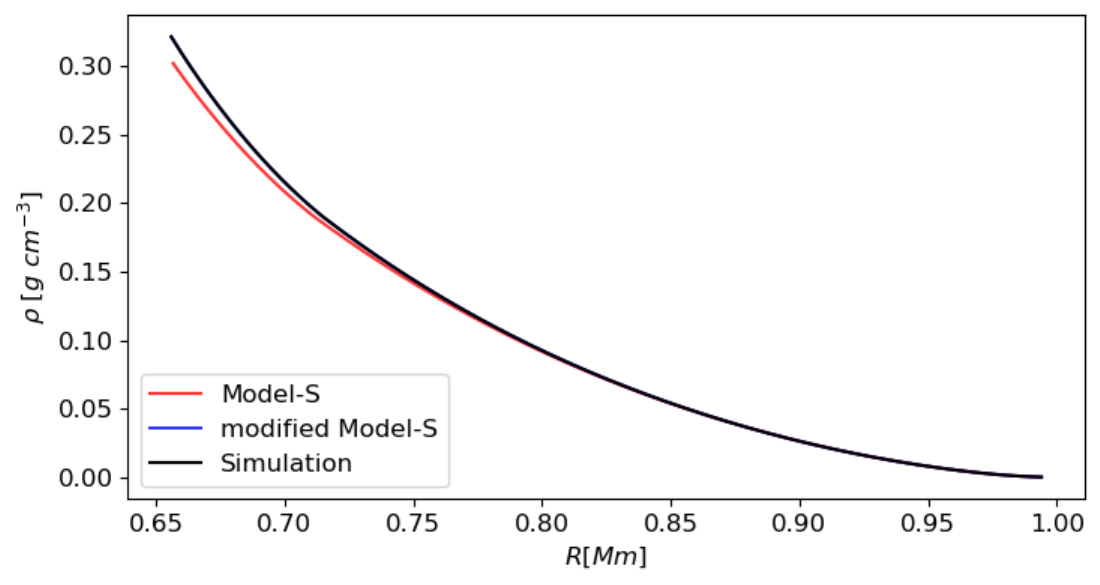}}
  \subfigure[Gas pressure]{\includegraphics[width=0.66\columnwidth]{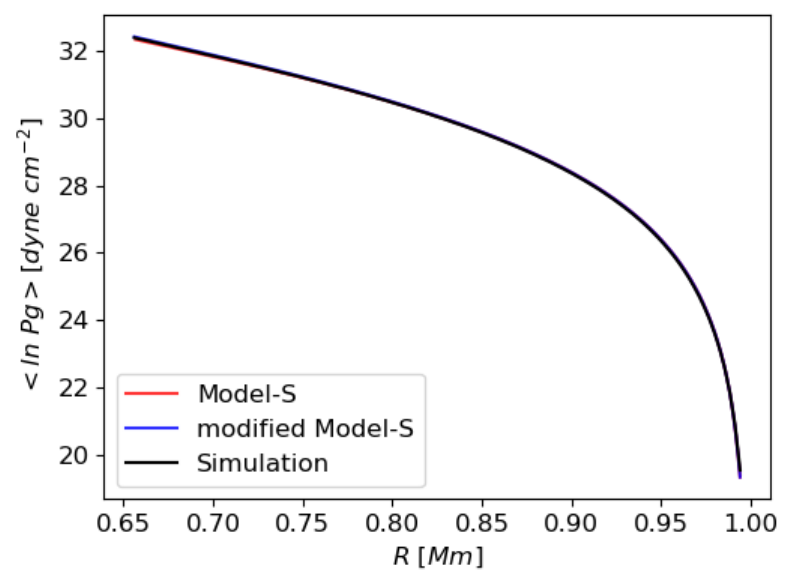}}
  \subfigure[Gas temperature]{\includegraphics[width=0.66\columnwidth]{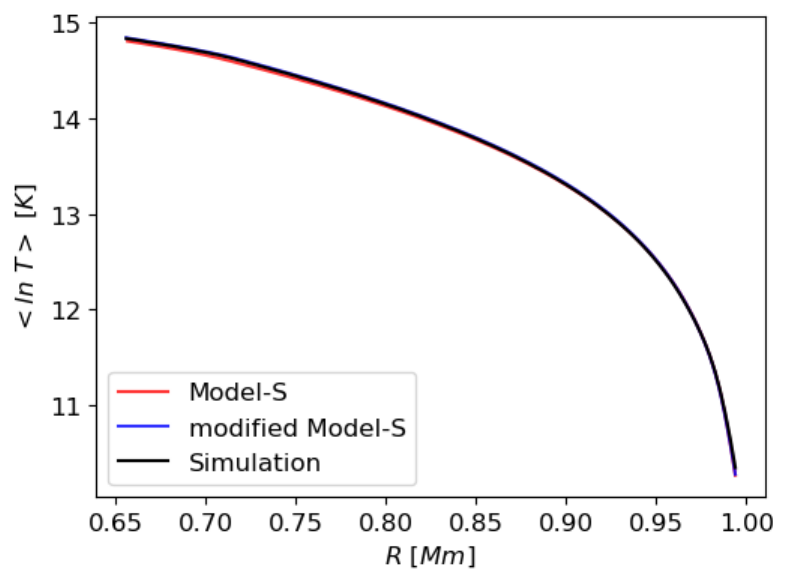}}
  \caption{Horizontally averaged quantities in the simulation (black), Model-S (red) with our EOS, and the Model-S, iterated for hydrostatic equilibrium using our EOS (blue). See text for details. }
  \label{fig:avg_prof}
\end{figure*}

\begin{figure}
\centering
    \includegraphics[width=0.85\columnwidth]{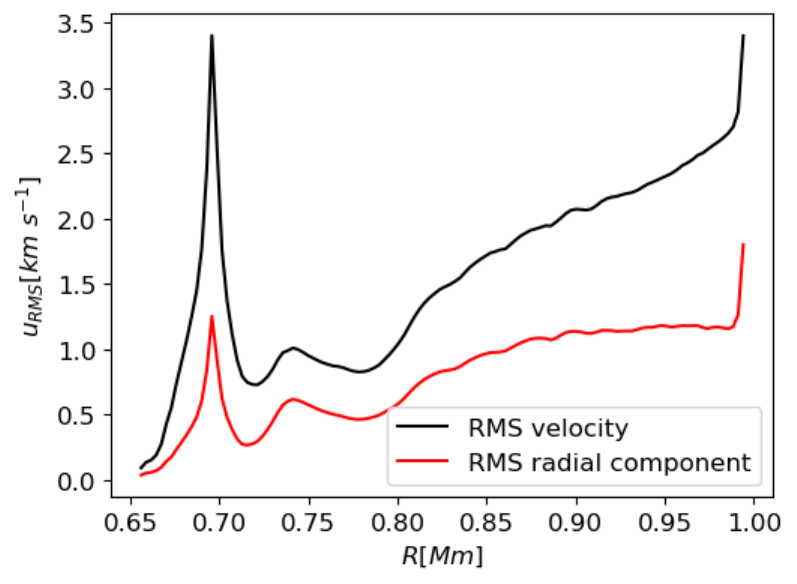}
    \caption{RMS velocity in our simulation. The massive spike in the deep part is at the convective-radiative boundary and is expected, while the simulation is relaxing.}
    \label{fig:rms_vel}
\end{figure}

In Figure \ref{fig:avg_prof} we show the horizontally averaged density, gas pressure and temperature (solid lines) with their extents (shaded areas) in our simulation compared to the Model-S as well as to the hydrostatic equilibrium, that we obtained by iterating the model-S with our EOS. The difference deep in the interior between the Model-S and hydrostatic equilibrium (modified Model-S) exemplifies the need to adjust the HSE for the EOS that is used. The visible differences are purely because of the use of different equation of state. The comparison between the simulation and the modified Model-S shows that the simulation maintains the equilibrium rather well.  We also show the RMS velocity in Figure \ref{fig:rms_vel}. The massive spike in the deep part is at the convective-radiative boundary and is expected while the simulation is relaxing and is expected to diminish with continued relaxation. Since the simulation has not been running for sufficient time for the meridional flows and differential rotation to settle into a steady-state, results for the large scale flow structures are yet inconclusive.

\section{Discussion}
\label{sec:discussion}
The new volleyball experiment setup provides a unique opportunity to study complex plasma dynamics, convection morphology, small and large scale magnetic field structures over a broad range of scales in the Sun, both at depth, in the surface layers, and possibly -- all the way to the corona.  

The simulations described in this paper were carried out on the Betzy supercomputer, located in Norway, and on one of the world’s largest supercomputers to date, LUMI in Finland. On LUMI, we ran the simulation using 96 nodes (each node contains 128 cores, totalling 12288 cores). With this size, in a 48 hour allocation period, the relaxation simulation evolves 48.3 hours, and thus gives evolution at essentially real solar-time.

The ideal allocation for the refined production run would be about 900 nodes ($\sim$ 110,000 cores). Unfortunately, LUMI normally limits job sizes to 512 nodes. However, we have executed a test run using 504 nodes, where the simulation reaches ~500 km above the solar surface. Details about the simulation are given in Appendix \ref{sec:photosphere}.

\subsection{Challenges}
With the number of Cartesian patches ranging from tens of thousands to millions, tessellation into a volleyball arrangement gives some challenges, both with respect to performance, and with respect to analysing and visualising the data.

The interpolations between tilted meshes are slightly costlier than in Cartesian mesh decomposition, but this can be significantly mitigated if the relative angles between adjacent patches are small enough to be handled as standard Cartesian coordinate interpolations (with only the fractional interpolation weights changing) to interpolate the guard zone data. This optimisation cannot, of course, be applied at the seams between the six volleyball `faces', but the affected volume fraction is very small, so the global cost is dominated by the optimised method.

We have in-house Python3 and Julia \citep{Bezanson_2014} language scripts to read in the snapshots and to process the data as a whole and as slices. The processed data can then be visualised with ParaView\footnote{ParaView is an open-source, multi-platform data analysis and visualisation application. \hyperlink{https://www.paraview.org/}{https://www.paraview.org/}}. Even without mesh refinement it is difficult to visualise the whole surface with a single, albeit powerful workstation\footnote{2$\times$ AMD EPYC 7742 64-Core Processor + 2$\times$ NVidia Tesla V100}, while with refined meshes one would either have to use the parallel capabilities of ParaView, or else zoom in to smaller regions, using the ``lazy reading'' capabilities of the low-level DISPATCH data access Python modules.  We are also continuously developing new tools and procedures to speed up the data processing.

The current snapshots, without mesh refinement, need 300 gigabytes of memory. A refined simulation, with e.g. 71 km resolution at the surface, would requires 2.3 terabytes of memory per snapshot. The storage requirements may increase even further, depending on refinement configuration or the need to extract additional, auxiliary, quantities. Now consider a future simulation with the core and photosphere included (see the test simulation and discussion of future work below), in the context of helioseismology. To do proper wave analysis we would require at least two weeks of solar time, with a 1 minute cadence for snapshots. That would correspond to 20160 snapshots, with approximately 50 petabytes of data, if the snapshots were stored ``as is''. Thus, for such studies we need to develop different data storage and retrieval solutions.

\subsection{Future work}

A natural next step will be continuing the simulation with a photosphere or lower chromosphere at the upper boundary. We wish to delay this step until the near future as adding a photospheric layer makes the simulations both more complicated and more expensive:
\begin{itemize}
    \item Simple surface cooling should be replaced by full multi-bin, short characteristics radiative heat transport;
    \item Spitzer conductivity becomes relevant in higher layers if we go to chromosphere and beyond;
    \item the resolution needed to sufficiently resolve the surface convective cells is of the order of a few tens of kilometre;
    \item upper boundary conditions need some modifications to minimise wave reflection;
    \item non-ideal effects (e.g.\ ambi-polar diffusion and time-dependent hydrogen, helium NLTE ionization) become important in the chromosphere and above;
    \item potential field extrapolation may be needed for magnetic field boundaries.
\end{itemize}
The current experiment includes the bottom of the convection zone, but excludes most of the radiative zone and the core. However, to do global helioseismological studies, the convectively stable central part of the Sun would need to be fully included in the simulations.  This will potentially require an addition of relativistic effects into the equation of state.  Including the central region of the Sun also requires switching to a normal Cartesian tessellation at some radius, to avoid the central singularity.  As with the seams already present, we do not expect this to lead to significant glitches.

The structural design of the experiment (composed of many small Cartesian grids) facilitates further exploration with zoom-in type of simulations  (regridding / refining in space and time). This technique is commonly used in simulations of star formation (e.g.\ \citealt{Kuffmeier_2017,Padoan_2020,Granzow_2026}). We will use this kind of setup to be able to focus on individual active regions without having to impose artificial initial and boundary conditions---these will instead be provided from the large scale simulation of the whole Sun.

To accommodate the need for high-cadence output for high resolution simulations (e.g. for helioseismology applications), we are developing an \textit{in-transit} data processing, which will produce high-cadence spherical or HEALPix datasets alongside the lower-cadence regular snapshots.

\section{Concluding remarks}
\label{sec:conclusions}
This paper describes a method to simulate the solar convective zone with a new `volleyball' mesh decomposition. The method provides a resolution that can be varied with radius, and while covering the full sphere still avoids coordinate singularities.

We show that the guard zone interpolations between patches do not lead to noticeable numerical artefacts, and that the transition between adjacent patches is generally smooth, also at the `seams' between the six volleyball faces. 

The simulations described here also demonstrate how the DISPATCH framework's flexibility allows us to gradually build up experiments by including more micro-physics, and by manipulating the mesh decomposition --- adding new layers, refining, or zooming in. This will be indispensable in the near future, when we have completed the relaxation and will be able to use the existing simulations as starting points for further experiments, with potentially different configurations and scientific goals. 

Variations on the experimental setup can be used to study, for example:
\begin{itemize}
    \item local and global dynamo effects
    \item active regions and flux emergence
    \item local and global helioseismology
    \item deep large-scale magnetic field structures
    \item overshoot regions at the base of the convection zone and consequential excitation of waves
    \item ...
\end{itemize}
Studies that will concentrate on specific features in space and/or time can use zoom-in techniques to focus on the areas of interest and enhance the realism (use additional physics modules) all the way to a similar complexity as e.g.\ \cite{Carlsson_2016}, with boundary conditions replaced by immersion in fully spherical initial models.

In a broader context, the volleyball setup can be applied with advantage to any problem with spherical geometry, without the need to handle coordinate singularities, and with local representations that always have one axis pointing in the vertical direction.  Contexts where this technique, in combination with the cost-saving advantages of local time-stepping, could significantly reduce simulation costs include for example simulating planet growth by pebble accretion (e.g. \citealt{Popovas_2018,Popovas_2019}, studying dynamics and escape of early planet atmospheres, pulsations of asymptotic giant branch stars, supernova explosions, as well as studies of accreting and interacting black holes.

\begin{acknowledgements}
This research was supported by the Research Council of Norway through its Centres of Excellence scheme, project number 262622, and through grants of computing time from the Programme for Supercomputing, as well as through the Synergy Grant number 810218 (ERC-2018-SyG) of the European Research Council, and Researcher Project for Early Career Scientists (FRIPRO) grant 359643 (``Resolving the Solar Convective Conundrum'', ReCon$^2$). We acknowledge the EuroHPC Joint Undertaking for awarding this project access to the EuroHPC supercomputer LUMI, hosted by CSC (Finland) and the LUMI consortium through a EuroHPC Extreme Scale Access call, project EHPC-EXT-2022E01-008. We also thank Jon P. Ramsey for the idea to tessellate the Cartesian patches on a 'Volleyball' surface.
\end{acknowledgements}

\bibliography{ms}

@ARTICLE{Amarsi_2016,
       author = {{Amarsi}, A.~M. and {Lind}, K. and {Asplund}, M. and {Barklem}, P.~S. and {Collet}, R.},
        title = "{Non-LTE line formation of Fe in late-type stars - III. 3D non-LTE analysis of metal-poor stars}",
      journal = {\mnras},
         year = 2016,
        month = dec,
       volume = {463},
       number = {2},
        pages = {1518-1533},
          doi = {10.1093/mnras/stw2077},
archivePrefix = {arXiv},
       eprint = {1608.06390},
 primaryClass = {astro-ph.SR},
       adsurl = {https://ui.adsabs.harvard.edu/abs/2016MNRAS.463.1518A}
}

@ARTICLE{Asplund_2021,
       author = {{Asplund}, M. and {Amarsi}, A.~M. and {Grevesse}, N.},
        title = "{The chemical make-up of the Sun: A 2020 vision}",
      journal = {\aap},
         year = 2021,
        month = sep,
       volume = {653},
          eid = {A141},
        pages = {A141},
          doi = {10.1051/0004-6361/202140445},
archivePrefix = {arXiv},
       eprint = {2105.01661},
 primaryClass = {astro-ph.SR},
       adsurl = {https://ui.adsabs.harvard.edu/abs/2021A&A...653A.141A}
}

@ARTICLE{Bezanson_2014,
       author = {{Bezanson}, Jeff and {Edelman}, Alan and {Karpinski}, Stefan and {Shah}, Viral B.},
        title = "{Julia: A Fresh Approach to Numerical Computing}",
      journal = {arXiv e-prints},
         year = 2014,
        month = nov,
          eid = {arXiv:1411.1607},
        pages = {arXiv:1411.1607},
archivePrefix = {arXiv},
       eprint = {1411.1607},
 primaryClass = {cs.MS},
       adsurl = {https://ui.adsabs.harvard.edu/abs/2014arXiv1411.1607B}
}

@ARTICLE{Brchnelova_2022,
       author = {{Brchnelova}, M. and {Zhang}, F. and {Leitner}, P. and {Perri}, B. and {Lani}, A. and {Poedts}, S.},
        title = "{Effects of mesh topology on MHD solution features in coronal simulations}",
      journal = {Journal of Plasma Physics},
         year = 2022,
        month = apr,
       volume = {88},
       number = {2},
          eid = {905880205},
        pages = {905880205},
          doi = {10.1017/S0022377822000241},
archivePrefix = {arXiv},
       eprint = {2202.13696},
 primaryClass = {astro-ph.SR},
       adsurl = {https://ui.adsabs.harvard.edu/abs/2022JPlPh..88b9005B}
}

@ARTICLE{Brun_2017,
       author = {{Brun}, Allan Sacha and {Strugarek}, Antoine and {Varela}, Jacobo and {Matt}, Sean P. and {Augustson}, Kyle C. and {Emeriau}, Constance and {DoCao}, Olivier Long and {Brown}, Benjamin and {Toomre}, Juri},
        title = "{On Differential Rotation and Overshooting in Solar-like Stars}",
      journal = {\apj},
         year = 2017,
        month = feb,
       volume = {836},
       number = {2},
          eid = {192},
        pages = {192},
          doi = {10.3847/1538-4357/aa5c40},
archivePrefix = {arXiv},
       eprint = {1702.06598},
 primaryClass = {astro-ph.SR},
       adsurl = {https://ui.adsabs.harvard.edu/abs/2017ApJ...836..192B}
}

@ARTICLE{Brun_2022,
       author = {{Brun}, Allan Sacha and {Strugarek}, Antoine and {Noraz}, Quentin and {Perri}, Barbara and {Varela}, Jacobo and {Augustson}, Kyle and {Charbonneau}, Paul and {Toomre}, Juri},
        title = "{Powering Stellar Magnetism: Energy Transfers in Cyclic Dynamos of Sun-like Stars}",
      journal = {\apj},
         year = 2022,
        month = feb,
       volume = {926},
       number = {1},
          eid = {21},
        pages = {21},
          doi = {10.3847/1538-4357/ac469b},
archivePrefix = {arXiv},
       eprint = {2201.13218},
 primaryClass = {astro-ph.SR},
       adsurl = {https://ui.adsabs.harvard.edu/abs/2022ApJ...926...21B}
}

@article{ Carlsson_2016,
	author = {{Carlsson}, Mats and {Hansteen}, Viggo H. and {Gudiksen}, Boris V. and {Leenaarts}, Jorrit and {De Pontieu}, Bart},
	title = {A publicly available simulation of an enhanced network region  of the Sun},
	DOI= "10.1051/0004-6361/201527226",
	url= "https://doi.org/10.1051/0004-6361/201527226",
	journal = {A\&A},
	year = 2016,
	volume = 585,
	pages = "A4",
	month = "",
}

@ARTICLE{Chen_2022,
       author = {{Chen}, Feng and {Rempel}, Matthias and {Fan}, Yuhong},
        title = "{A Comprehensive Radiative Magnetohydrodynamics Simulation of Active Region Scale Flux Emergence from the Convection Zone to the Corona}",
      journal = {\apj},
         year = 2022,
        month = oct,
       volume = {937},
       number = {2},
          eid = {91},
        pages = {91},
          doi = {10.3847/1538-4357/ac8f95},
archivePrefix = {arXiv},
       eprint = {2106.14055},
 primaryClass = {astro-ph.SR},
       adsurl = {https://ui.adsabs.harvard.edu/abs/2022ApJ...937...91C}
}

@article{Chen_fc_2022,
author = {Chen, Siyuan and Peng, Xindong and Chang, Jun and Kang, Yanzhen and Gu, Juan},
title = {Positivity-preserving transport with local and global mass conservation on Yin–Yang grids},
journal = {Quarterly Journal of the Royal Meteorological Society},
volume = {n/a},
year = 2026,
month = June,
number = {n/a},
pages = {e70246},
doi = {https://doi.org/10.1002/qj.70246},
url = {https://rmets.onlinelibrary.wiley.com/doi/abs/10.1002/qj.70246},
eprint = {https://rmets.onlinelibrary.wiley.com/doi/pdf/10.1002/qj.70246},
note = {e70246 7111199}
}

@ARTICLE{Freytag_2017,
       author = {{Freytag}, B. and {Liljegren}, S. and {H{\"o}fner}, S.},
        title = "{Global 3D radiation-hydrodynamics models of AGB stars. Effects of convection and radial pulsations on atmospheric structures}",
      journal = {\aap},
         year = 2017,
        month = apr,
       volume = {600},
          eid = {A137},
        pages = {A137},
          doi = {10.1051/0004-6361/201629594},
archivePrefix = {arXiv},
       eprint = {1702.05433},
 primaryClass = {astro-ph.SR},
       adsurl = {https://ui.adsabs.harvard.edu/abs/2017A&A...600A.137F}
}

@ARTICLE{Goldberg_2022,
       author = {{Goldberg}, Jared A. and {Jiang}, Yan-Fei and {Bildsten}, Lars},
        title = "{Numerical Simulations of Convective Three-dimensional Red Supergiant Envelopes}",
      journal = {\apj},
         year = 2022,
        month = apr,
       volume = {929},
       number = {2},
          eid = {156},
        pages = {156},
          doi = {10.3847/1538-4357/ac5ab3},
archivePrefix = {arXiv},
       eprint = {2110.03261},
 primaryClass = {astro-ph.SR},
       adsurl = {https://ui.adsabs.harvard.edu/abs/2022ApJ...929..156G}
}

@ARTICLE{Granzow_2026,
       author = {{Granzow Holm}, C. and {Lambrechts}, M. and {Kuffmeier}, M. and {Johansen}, A. and {Haugb{\o}lle}, T. and {Nordlund}, {\r{A}}.},
        title = "{Accretion across scales: Streamers, surface-layer transport, and rapid replenishment in young protoplanetary discs}",
      journal = {\aap},
         year = 2026,
        month = oct,
       volume = {714},
          eid = {A161},
        pages = {A161},
          doi = {10.1051/0004-6361/202659974},
archivePrefix = {arXiv},
       eprint = {2608.02204},
 primaryClass = {astro-ph.EP},
       adsurl = {https://ui.adsabs.harvard.edu/abs/2026A&A...714A.161G}
}

@article{
JCD_1996,
author = {J. Christensen-Dalsgaard  and W. Däppen  and S. V. Ajukov  and E. R. Anderson  and H. M. Antia  and S. Basu  and V. A. Baturin  and G. Berthomieu  and B. Chaboyer  and S. M. Chitre  and A. N. Cox  and P. Demarque  and J. Donatowicz  and W. A. Dziembowski  and M. Gabriel  and D. O. Gough  and D. B. Guenther  and J. A. Guzik  and J. W. Harvey  and F. Hill  and G. Houdek  and C. A. Iglesias  and A. G. Kosovichev  and J. W. Leibacher  and P. Morel  and C. R. Proffitt  and J. Provost  and J. Reiter  and E. J. Rhodes  and F. J. Rogers  and I. W. Roxburgh  and M. J. Thompson  and R. K. Ulrich },
title = {The Current State of Solar Modeling},
journal = {Science},
volume = {272},
number = {5266},
pages = {1286-1292},
year = {1996},
doi = {10.1126/science.272.5266.1286},
URL = {https://www.science.org/doi/abs/10.1126/science.272.5266.1286},
eprint = {https://www.science.org/doi/pdf/10.1126/science.272.5266.1286}}

@ARTICLE{Druett_2022,
       author = {{Druett}, M.~K. and {Leenaarts}, J. and {Carlsson}, M. and {Szydlarski}, M.},
        title = "{Formation and heating of chromospheric fibrils in a radiation-MHD simulation}",
      journal = {\aap},
         year = 2022,
        month = sep,
       volume = {665},
          eid = {A6},
        pages = {A6},
          doi = {10.1051/0004-6361/202142399},
archivePrefix = {arXiv},
       eprint = {2112.08245},
 primaryClass = {astro-ph.SR},
       adsurl = {https://ui.adsabs.harvard.edu/abs/2022A&A...665A...6D}
}

@article{Eitner_2025,
   title={M3DIS – A grid of 3D radiation-hydrodynamics stellar atmosphere models for stellar surveys: II. Carbon-enhanced metal-poor stars},
   volume={703},
   ISSN={1432-0746},
   url={http://dx.doi.org/10.1051/0004-6361/202555406},
   DOI={10.1051/0004-6361/202555406},
   journal={\aap},
   publisher={EDP Sciences},
   author={Eitner, Philipp and Bergemann, Maria and Hoppe, Richard and Storm, Nicholas and Lipatova, Veronika and Glover, Simon C. O. and Klessen, Ralf S. and Nordlund, Åke and Popovas, Andrius},
   year={2025},
   month=Nov, pages={A199} }

@ARTICLE{Finley_2022,
       author = {{Finley}, A.~J. and {Brun}, A.~S. and {Carlsson}, M. and {Szydlarski}, M. and {Hansteen}, V. and {Shoda}, M.},
        title = "{Stirring the base of the solar wind: On heat transfer and vortex formation}",
      journal = {\aap},
         year = 2022,
        month = sep,
       volume = {665},
          eid = {A118},
        pages = {A118},
          doi = {10.1051/0004-6361/202243947},
archivePrefix = {arXiv},
       eprint = {2207.02878},
 primaryClass = {astro-ph.SR},
       adsurl = {https://ui.adsabs.harvard.edu/abs/2022A&A...665A.118F}
}

@ARTICLE{Guerrero_2022,
       author = {{Guerrero}, G. and {Stejko}, A.~M. and {Kosovichev}, A.~G. and {Smolarkiewicz}, P. K and {Strugarek}, A.},
        title = "{Implicit large eddy simulations of global solar convection: effects of numerical resolution in non-rotating and rotating cases}",
      journal = {arXiv e-prints},
         year = 2022,
        month = aug,
          eid = {arXiv:2208.05738},
        pages = {arXiv:2208.05738},
archivePrefix = {arXiv},
       eprint = {2208.05738},
 primaryClass = {astro-ph.SR},
       adsurl = {https://ui.adsabs.harvard.edu/abs/2022arXiv220805738G}
}

@ARTICLE{Hotta_2012,
       author = {{Hotta}, H. and {Rempel}, M. and {Yokoyama}, T. and {Iida}, Y. and {Fan}, Y.},
        title = "{Numerical calculation of convection with reduced speed of sound technique}",
      journal = {\aap},
         year = 2012,
        month = mar,
       volume = {539},
          eid = {A30},
        pages = {A30},
          doi = {10.1051/0004-6361/201118268},
archivePrefix = {arXiv},
       eprint = {1201.1061},
 primaryClass = {astro-ph.SR},
       adsurl = {https://ui.adsabs.harvard.edu/abs/2012A&A...539A..30H}
}

@ARTICLE{Hotta_2014,
       author = {{Hotta}, H. and {Rempel}, M. and {Yokoyama}, T.},
        title = "{High-resolution Calculations of the Solar Global Convection with the Reduced Speed of Sound Technique. I. The Structure of the Convection and the Magnetic Field without the Rotation}",
      journal = {\apj},
         year = 2014,
        month = may,
       volume = {786},
       number = {1},
          eid = {24},
        pages = {24},
          doi = {10.1088/0004-637X/786/1/24},
archivePrefix = {arXiv},
       eprint = {1402.5008},
 primaryClass = {astro-ph.SR},
       adsurl = {https://ui.adsabs.harvard.edu/abs/2014ApJ...786...24H}
}

@ARTICLE{Hotta_2015,
       author = {{Hotta}, H. and {Rempel}, M. and {Yokoyama}, T.},
        title = "{High-resolution Calculation of the Solar Global Convection with the Reduced Speed of Sound Technique. II. Near Surface Shear Layer with the Rotation}",
      journal = {\apj},
         year = 2015,
        month = jan,
       volume = {798},
       number = {1},
          eid = {51},
        pages = {51},
          doi = {10.1088/0004-637X/798/1/51},
archivePrefix = {arXiv},
       eprint = {1410.7093},
 primaryClass = {astro-ph.SR},
       adsurl = {https://ui.adsabs.harvard.edu/abs/2015ApJ...798...51H}
}

@ARTICLE{Hotta_2019,
       author = {{Hotta}, H. and {Iijima}, H. and {Kusano}, K.},
        title = "{Weak influence of near-surface layer on solar deep convection zone revealed by comprehensive simulation from base to surface}",
      journal = {Science Advances},
         year = 2019,
        month = jan,
       volume = {5},
       number = {1},
        pages = {2307},
          doi = {10.1126/sciadv.aau2307},
       adsurl = {https://ui.adsabs.harvard.edu/abs/2019SciA....5.2307H}
}

@ARTICLE{Hotta_2020,
       author = {{Hotta}, H. and {Iijima}, H.},
        title = "{On rising magnetic flux tube and formation of sunspots in a deep domain}",
      journal = {\mnras},
         year = 2020,
        month = may,
       volume = {494},
       number = {2},
        pages = {2523-2537},
          doi = {10.1093/mnras/staa844},
archivePrefix = {arXiv},
       eprint = {2003.10583},
 primaryClass = {astro-ph.SR},
       adsurl = {https://ui.adsabs.harvard.edu/abs/2020MNRAS.494.2523H}
}

@ARTICLE{Hotta_2021,
       author = {{Hotta}, H. and {Kusano}, K.},
        title = "{Solar differential rotation reproduced with high-resolution simulation}",
      journal = {Nature Astronomy},
         year = 2021,
        month = sep,
       volume = {5},
        pages = {1100-1102},
          doi = {10.1038/s41550-021-01459-0},
archivePrefix = {arXiv},
       eprint = {2109.06280},
 primaryClass = {astro-ph.SR},
       adsurl = {https://ui.adsabs.harvard.edu/abs/2021NatAs...5.1100H}
}

@ARTICLE{Hotta_2022,
       author = {{Hotta}, H. and {Kusano}, K. and {Shimada}, R.},
        title = "{Generation of Solar-like Differential Rotation}",
      journal = {\apj},
         year = 2022,
        month = jul,
       volume = {933},
       number = {2},
          eid = {199},
        pages = {199},
          doi = {10.3847/1538-4357/ac7395},
archivePrefix = {arXiv},
       eprint = {2202.04183},
 primaryClass = {astro-ph.SR},
       adsurl = {https://ui.adsabs.harvard.edu/abs/2022ApJ...933..199H}
}

@software{Irwin_2012,
       author = {{Irwin}, Alan W.},
        title = "{FreeEOS: Equation of State for stellar interiors calculations}",
 howpublished = {Astrophysics Source Code Library, record ascl:1211.002},
         year = 2012,
        month = nov,
          eid = {ascl:1211.002},
archivePrefix = {ascl},
       eprint = {1211.002},
       adsurl = {https://ui.adsabs.harvard.edu/abs/2012ascl.soft11002I}
}

@article{Kageyama_2003,
author = {Kageyama, Akira and Sato, Tetsuya},
title = {“Yin-Yang grid”: An overset grid in spherical geometry},
journal = {Geochemistry, Geophysics, Geosystems},
volume = {5},
number = {9},
pages = {},
doi = {https://doi.org/10.1029/2004GC000734},
url = {https://agupubs.onlinelibrary.wiley.com/doi/abs/10.1029/2004GC000734},
year = {2004}
}

@ARTICLE{Kapyla_2021,
       author = {{K{\"a}pyl{\"a}}, P.~J.},
        title = "{Star-in-a-box simulations of fully convective stars}",
      journal = {\aap},
         year = 2021,
        month = jul,
       volume = {651},
          eid = {A66},
        pages = {A66},
          doi = {10.1051/0004-6361/202040049},
archivePrefix = {arXiv},
       eprint = {2012.01259},
 primaryClass = {astro-ph.SR},
       adsurl = {https://ui.adsabs.harvard.edu/abs/2021A&A...651A..66K}
}

@ARTICLE{Kapyla_2025,
       author = {{K{\"a}pyl{\"a}}, P.~J.},
        title = "{Simulations of entropy rain-driven convection}",
      journal = {\aap},
         year = 2025,
        month = jun,
       volume = {698},
          eid = {L13},
        pages = {L13},
          doi = {10.1051/0004-6361/202554952},
archivePrefix = {arXiv},
       eprint = {2504.00738},
 primaryClass = {astro-ph.SR},
       adsurl = {https://ui.adsabs.harvard.edu/abs/2025A&A...698L..13K}
}

@ARTICLE{Kohutova_2021,
       author = {{Kohutova}, P. and {Popovas}, A.},
        title = "{Excitation and evolution of coronal oscillations in self-consistent 3D radiative MHD simulations of the solar atmosphere}",
      journal = {\aap},
         year = 2021,
        month = mar,
       volume = {647},
          eid = {A81},
        pages = {A81},
          doi = {10.1051/0004-6361/202039491},
archivePrefix = {arXiv},
       eprint = {2101.06430},
 primaryClass = {astro-ph.SR},
       adsurl = {https://ui.adsabs.harvard.edu/abs/2021A&A...647A..81K}
}

@ARTICLE{Kuffmeier_2017,
       author = {{Kuffmeier}, Michael and {Haugb{\o}lle}, Troels and {Nordlund}, {\r{A}}ke},
        title = "{Zoom-in Simulations of Protoplanetary Disks Starting from GMC Scales}",
      journal = {\apj},
         year = 2017,
        month = sep,
       volume = {846},
       number = {1},
          eid = {7},
        pages = {7},
          doi = {10.3847/1538-4357/aa7c64},
archivePrefix = {arXiv},
       eprint = {1611.10360},
 primaryClass = {astro-ph.SR},
       adsurl = {https://ui.adsabs.harvard.edu/abs/2017ApJ...846....7K}
}

@ARTICLE{Larson_2018,
       author = {{Larson}, Timothy P. and {Schou}, Jesper},
        title = "{Global-Mode Analysis of Full-Disk Data from the Michelson Doppler Imager and the Helioseismic and Magnetic Imager}",
      journal = {\solphys},
         year = 2018,
        month = jan,
       volume = {293},
       number = {2},
          eid = {29},
        pages = {29},
          doi = {10.1007/s11207-017-1201-5},
       adsurl = {https://ui.adsabs.harvard.edu/abs/2018SoPh..293...29L}
}

@ARTICLE{Miesch_2008,
       author = {{Miesch}, Mark S. and {Brun}, Allan Sacha and {DeRosa}, Marc L. and {Toomre}, Juri},
        title = "{Structure and Evolution of Giant Cells in Global Models of Solar Convection}",
      journal = {\apj},
         year = 2008,
        month = jan,
       volume = {673},
       number = {1},
        pages = {557-575},
          doi = {10.1086/523838},
archivePrefix = {arXiv},
       eprint = {0707.1460},
 primaryClass = {astro-ph},
       adsurl = {https://ui.adsabs.harvard.edu/abs/2008ApJ...673..557M}
}

@ARTICLE{Nordlund_2009,
       author = {{Nordlund}, {\r{A}}ke and {Stein}, Robert F. and {Asplund}, Martin},
        title = "{Solar Surface Convection}",
      journal = {Living Reviews in Solar Physics},
         year = 2009,
        month = dec,
       volume = {6},
       number = {1},
          eid = {2},
        pages = {2},
          doi = {10.12942/lrsp-2009-2},
       adsurl = {https://ui.adsabs.harvard.edu/abs/2009LRSP....6....2N}
}

@ARTICLE{Nordlund_2018MNRAS,
       author = {{Nordlund}, {\r{A}}ke and {Ramsey}, Jon P. and {Popovas}, Andrius and {K{\"u}ffmeier}, Michael},
        title = "{DISPATCH: a numerical simulation framework for the exa-scale era - I. Fundamentals}",
      journal = {\mnras},
         year = 2018,
        month = jun,
       volume = {477},
       number = {1},
        pages = {624-638},
          doi = {10.1093/mnras/sty599},
archivePrefix = {arXiv},
       eprint = {1705.10774},
 primaryClass = {astro-ph.IM},
       adsurl = {https://ui.adsabs.harvard.edu/abs/2018MNRAS.477..624N}
}

@ARTICLE{Padoan_2020,
       author = {{Padoan}, Paolo and {Pan}, Liubin and {Juvela}, Mika and {Haugb{\o}lle}, Troels and {Nordlund}, {\r{A}}ke},
        title = "{The Origin of Massive Stars: The Inertial-inflow Model}",
      journal = {\apj},
         year = 2020,
        month = sep,
       volume = {900},
       number = {1},
          eid = {82},
        pages = {82},
          doi = {10.3847/1538-4357/abaa47},
archivePrefix = {arXiv},
       eprint = {1911.04465},
 primaryClass = {astro-ph.GA},
       adsurl = {https://ui.adsabs.harvard.edu/abs/2020ApJ...900...82P}
}

@ARTICLE{Parenti_2014,
       author = {{Parenti}, Susanna},
        title = "{Solar Prominences: Observations}",
      journal = {Living Reviews in Solar Physics},
         year = 2014,
        month = mar,
       volume = {11},
       number = {1},
          eid = {1},
        pages = {1},
          doi = {10.12942/lrsp-2014-1},
       adsurl = {https://ui.adsabs.harvard.edu/abs/2014LRSP...11....1P}
}

@article{Peng_2006,
author = {Peng, Xindong and Xiao, Feng and Takahashi, Keiko},
title = {Conservative constraint for a quasi-uniform overset grid on the sphere},
journal = {Quarterly Journal of the Royal Meteorological Society},
volume = {132},
number = {616},
pages = {979-996},
doi = {https://doi.org/10.1256/qj.05.18},
url = {https://rmets.onlinelibrary.wiley.com/doi/abs/10.1256/qj.05.18},
eprint = {https://rmets.onlinelibrary.wiley.com/doi/pdf/10.1256/qj.05.18},
year = {2006}
}

@ARTICLE{Popovas_2025a,
       author = {{Popovas}, Andrius},
        title = "{DISPATCH methods: An approximate, entropy-based Riemann solver for ideal magnetohydrodynamics}",
      journal = {\aap},
         year = 2025,
        month = jun,
       volume = {698},
          eid = {A69},
        pages = {A69},
          doi = {10.1051/0004-6361/202554028},
archivePrefix = {arXiv},
       eprint = {2211.02438},
 primaryClass = {astro-ph.IM},
       adsurl = {https://ui.adsabs.harvard.edu/abs/2025A&A...698A..69P}
}

@ARTICLE{Popovas_2018,
       author = {{Popovas}, Andrius and {Nordlund}, {\r{A}}ke and {Ramsey}, Jon P. and {Ormel}, Chris W.},
        title = "{Pebble dynamics and accretion on to rocky planets - I. Adiabatic and convective models}",
      journal = {\mnras},
         year = 2018,
        month = oct,
       volume = {479},
       number = {4},
        pages = {5136-5156},
          doi = {10.1093/mnras/sty1752},
archivePrefix = {arXiv},
       eprint = {1801.07707},
 primaryClass = {astro-ph.EP},
       adsurl = {https://ui.adsabs.harvard.edu/abs/2018MNRAS.479.5136P}
}

@ARTICLE{Popovas_2019,
       author = {{Popovas}, Andrius and {Nordlund}, {\r{A}}ke and {Ramsey}, Jon P.},
        title = "{Pebble dynamics and accretion on to rocky planets - II. Radiative models}",
      journal = {\mnras},
         year = 2019,
        month = jan,
       volume = {482},
       number = {1},
        pages = {L107-L111},
          doi = {10.1093/mnrasl/sly197},
archivePrefix = {arXiv},
       eprint = {1810.07048},
 primaryClass = {astro-ph.EP},
       adsurl = {https://ui.adsabs.harvard.edu/abs/2019MNRAS.482L.107P}
}

@ARTICLE{Przybylski_2022,
       author = {{Przybylski}, D. and {Cameron}, R. and {Solanki}, S.~K. and {Rempel}, M. and {Leenaarts}, J. and {Anusha}, L.~S. and {Witzke}, V. and {Shapiro}, A.~I.},
        title = "{Chromospheric extension of the MURaM code}",
      journal = {\aap},
         year = 2022,
        month = aug,
       volume = {664},
          eid = {A91},
        pages = {A91},
          doi = {10.1051/0004-6361/202141230},
archivePrefix = {arXiv},
       eprint = {2204.03126},
 primaryClass = {astro-ph.SR},
       adsurl = {https://ui.adsabs.harvard.edu/abs/2022A&A...664A..91P}
}

@article{Ramsey_2018,
doi = {10.1088/1742-6596/1031/1/012021},
url = {https://dx.doi.org/10.1088/1742-6596/1031/1/012021},
year = {2018},
month = {may},
publisher = {IOP Publishing},
volume = {1031},
number = {1},
pages = {012021},
author = {Ramsey, J P and Haugbølle, T and Nordlund, \r{A}},
title = {A simple and efficient solver for self-gravity in the DISPATCH astrophysical simulation framework},
journal = {Journal of Physics: Conference Series}
}

@ARTICLE{Reale_2010,
       author = {{Reale}, Fabio},
        title = "{Coronal Loops: Observations and Modeling of Confined Plasma}",
      journal = {Living Reviews in Solar Physics},
         year = 2010,
        month = nov,
       volume = {7},
       number = {1},
          eid = {5},
        pages = {5},
          doi = {10.12942/lrsp-2010-5},
archivePrefix = {arXiv},
       eprint = {1010.5927},
 primaryClass = {astro-ph.SR},
       adsurl = {https://ui.adsabs.harvard.edu/abs/2010LRSP....7....5R}
}

@ARTICLE{Rieutord_2010,
       author = {{Rieutord}, Michel and {Rincon}, Fran{\c{c}}ois},
        title = "{The Sun's Supergranulation}",
      journal = {Living Reviews in Solar Physics},
         year = 2010,
        month = dec,
       volume = {7},
       number = {1},
          eid = {2},
        pages = {2},
          doi = {10.12942/lrsp-2010-2},
archivePrefix = {arXiv},
       eprint = {1005.5376},
 primaryClass = {astro-ph.SR},
       adsurl = {https://ui.adsabs.harvard.edu/abs/2010LRSP....7....2R}
}

@ARTICLE{Ronchi_1996,
       author = {{Ronchi}, C. and {Iacono}, R. and {Paolucci}, P.~S.},
        title = "{The ``Cubed Sphere'': A New Method for the Solution of Partial Differential Equations in Spherical Geometry}",
      journal = {Journal of Computational Physics},
         year = 1996,
        month = mar,
       volume = {124},
       number = {1},
        pages = {93-114},
          doi = {10.1006/jcph.1996.0047},
       adsurl = {https://ui.adsabs.harvard.edu/abs/1996JCoPh.124...93R}
}

@ARTICLE{Stein_2012,
       author = {{Stein}, Robert F. and {Nordlund}, {\r{A}}ke},
        title = "{On the Formation of Active Regions}",
      journal = {\apjl},
         year = 2012,
        month = jul,
       volume = {753},
       number = {1},
          eid = {L13},
        pages = {L13},
          doi = {10.1088/2041-8205/753/1/L13},
archivePrefix = {arXiv},
       eprint = {1207.4248},
 primaryClass = {astro-ph.SR},
       adsurl = {https://ui.adsabs.harvard.edu/abs/2012ApJ...753L..13S}
}
\begin{appendix}

\section{Surface cooling calibration}
\label{sec:surf_cool_calib}

In the Pilot experiment the upper boundary lies at 0.998 R$_\odot$, $\approx$1.4 Mm below the optical surface. The radiative losses that drive the convection are therefore replaced by the parametrised cooling of Sect. \ref{subsubsec:surface_cooling}. This appendix describes how its four parameters were calibrated, in order for the parametrised cooling to reproduce four properties of the cold material produced by radiative losses at the surface:
\begin{enumerate}
    \item its entropy, which sets the plume contrast;
    \item its velocity, which sets the enthalpy flux;
    \item the timescale on which it forms, which determines whether the convection granules split as they expand, and hence,
    \item the horizontal scale at which energy is deposited.
\end{enumerate}

Two parameters $h_b$ and $\psi$ are fixed a priori (see below). The other two, $S_{\rm target}$ and $\tau_{\rm top}$, are calibrated against a reference simulation. The reference simulation has the same physics and near surface resolution as the Pilot, set up in the `sandbox` (Cartesian) configuration; comparing like with like isolates the effect of the cooling prescription. The models extend above the optical surface and the reference model computes the radiative cooling explicitly, using ray-based radiative transfer with multi-bin opacities. $S_{\rm target}$ is measured directly from the reference, $\tau_{\rm top}$ is chosen by scoring a scan of runs against it.

At R=0.998 R$_\odot$ the Rosseland mean opacity from the opacity tables gives $\kappa\bar\rho \approx 1.5\times10^{-2}$~cm$^{-1}$ at $\bar\rho \approx 4.9\times10^{-6}$~g~cm$^{-3}$, thus the optical depth across one 100 km cell is $\sim\!10^{5}$. Thus $h_b$ has to be as thin as simulations allow; i.e., cooling should be essentially constrained to a single cell thickness. Closer to the actual surface, especially in higher resolution simulations $h_b$ may become larger than $\Delta s$. In such cases $h_b$ is somewhat degenerate with $\tau_{\rm top}$ and a grid of calibration models is needed.

We take S$_{\rm floor}$ from the 3D reference rather than from a 1D solar model. A 1D model provides only the horizontally averaged entropy, whereas $_{\rm floor}$ must represent the coldest descending material, whose deviation from the mean the 1D model does not contain.

\subsection{Step 1 --- the reference entropy floor \texorpdfstring{$S_{\rm target}$}{Starget}}

A quantity of interest is the entropy of the coldest descending material. We use the median of a low percentile (0.5\%) of the specific entropy distribution, computed over a range of snapshots (around 11 solar hours). The median removes the oscillatory effects from $p$-mode displacements of the  horizontal plane.

\begin{figure}
\centering
    \includegraphics[width=0.99\columnwidth]{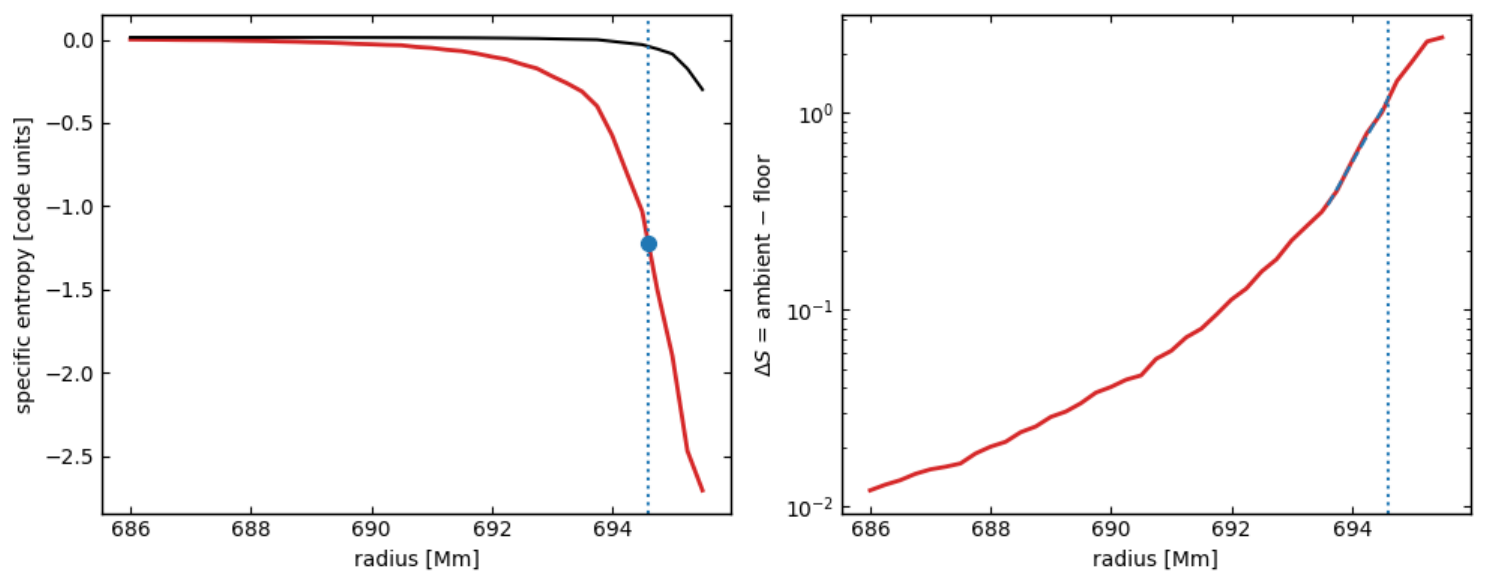}
    \caption{Reference entropy calibration. Left: floor at the truncation depth (R=0.998 R$_\odot$). The black curve is the horizontally averaged entropy profile. Right: contrast and its local decay length.}
    \label{fig:ss_floor}
\end{figure}

Figure~\ref{fig:ss_floor} shows the target entropy of the cooling, $S_{\rm target}(z)$, as a function of depth. The floor is strongly depth dependent: the contrast between the coldest material and the horizontal median decays with an e-folding length of about 1~Mm immediately below the surface, increasing to roughly 3~Mm some 10~Mm deeper. This decay is expected, albeit strongly resolution-dependent --- descending plumes are heated by compression, the cold lanes narrow, and become diluted by mixing. A calibration performed for one upper boundary radius is not directly applicable at a different radius, and the target must be re-calibrated for any given upper boundary radius. The adopted parameter values are valid for the code units, configuration, and equation-of-state described in Section \ref{sec:experiment_design} and may not be applicable if for example the entropy null-point is different in the EOS. 

\subsection{Step 2 --- the cooling time scale \texorpdfstring{$\tau_{\rm top}$}{tautop}}

$\tau_{\rm top}$ is physically bound from both sides. The granules split (`explode') when a cold downflow forms inside an expanding cell, which requires the cooling to be fast enough to establish a low-entropy patch before the granule advects it to the cell boundary. Beyond the granule lifetime no interior downflow forms, cells survive until destroyed by interaction with neighbours, and the size distribution shifts to larger scales. 

The theoretical radiative cooling timescale can be derived from
\begin{equation}
\label{eq:tau_estimate}
    \tau_{\rm top} \simeq \frac{\rho\, c_v\, T\, \ell}{\sigma_{\rm B}T^4},
\end{equation}
with $\rho \sim 10^{-7}$~g~cm$^{-3}$, $T \sim 6000$~K, $c_v \sim 10^{8} - 10^{9}$~erg~g$^{-1}$~K$^{-1}$ and $\ell \sim 100$~km, giving a $\tau_{\rm rad}$ of order of 10 -- 100 seconds. Deeper down in the interior such an estimate is not straightforward. There, our aim is to construct a pseudo-surface, which will deviate somewhat from the initial profile, but result in a convection morphology comparable to the surface simulations. 

Note that the surface cooling acts on every cell in the plane, regardless of whether it lies in an upflow or a downflow. If $\tau_{\rm top}$ is too short compared with the residence time in the cooling layer, every cell reaches the target and the boundary becomes a uniformly cold surface instead of one with intermittent cold lanes. 

\begin{figure*}
  \centering
  \resizebox{\hsize}{!}
  {\includegraphics[width=0.99\columnwidth]{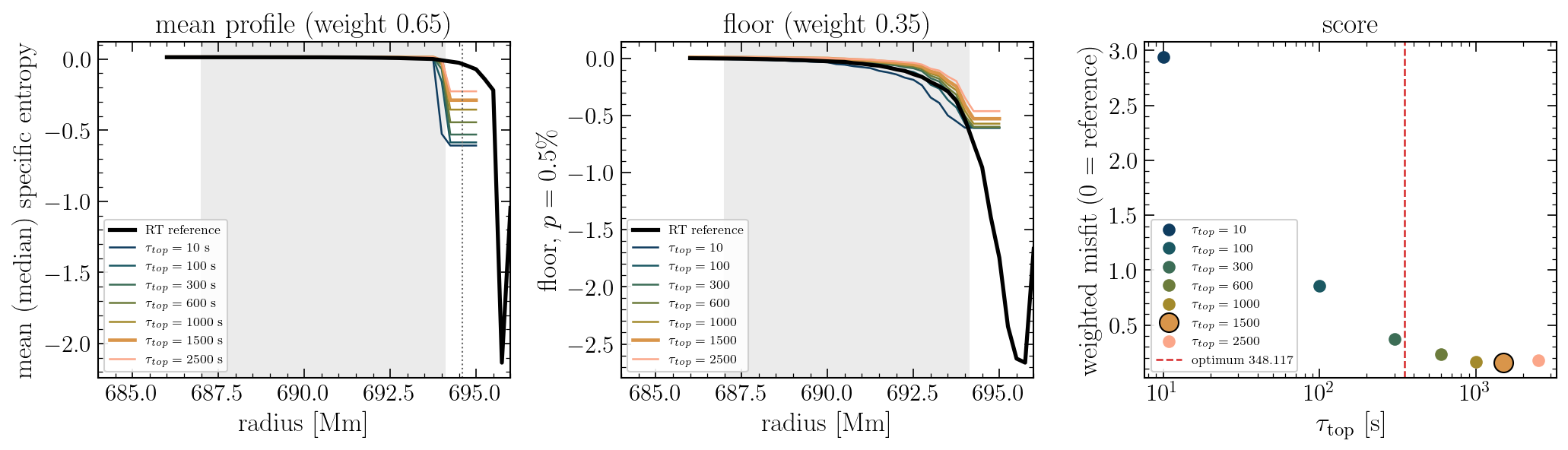}}
  \caption{Surface cooling timescale $\tau_{\rm top}$ calibration. Left: Comparison of horizontal mean entropy profiles. Middle: Comparison of the floor entropy values. Right: Scoring to obtain the best fitting $\tau_{\rm top}$. The red dashed vertical line denotes the most likely optimum value for this test.}
  \label{fig:ss_h_calib}
\end{figure*}

Three diagnostics are used, which identify discrepancies in distinct ways. The first is the horizontally averaged entropy profile, with the entropy floor below the truncation radius, compared with the reference using the same percentile statistic. The difference from the reference model is compared and scored, as illustrated in Figure \ref{fig:ss_h_calib}. Too strong cooling lowers the average entropy profile, while too weak cooling will result in too low contrast.

\begin{figure*}
  \centering
  \resizebox{\hsize}{!}
  {\includegraphics[width=0.99\columnwidth]{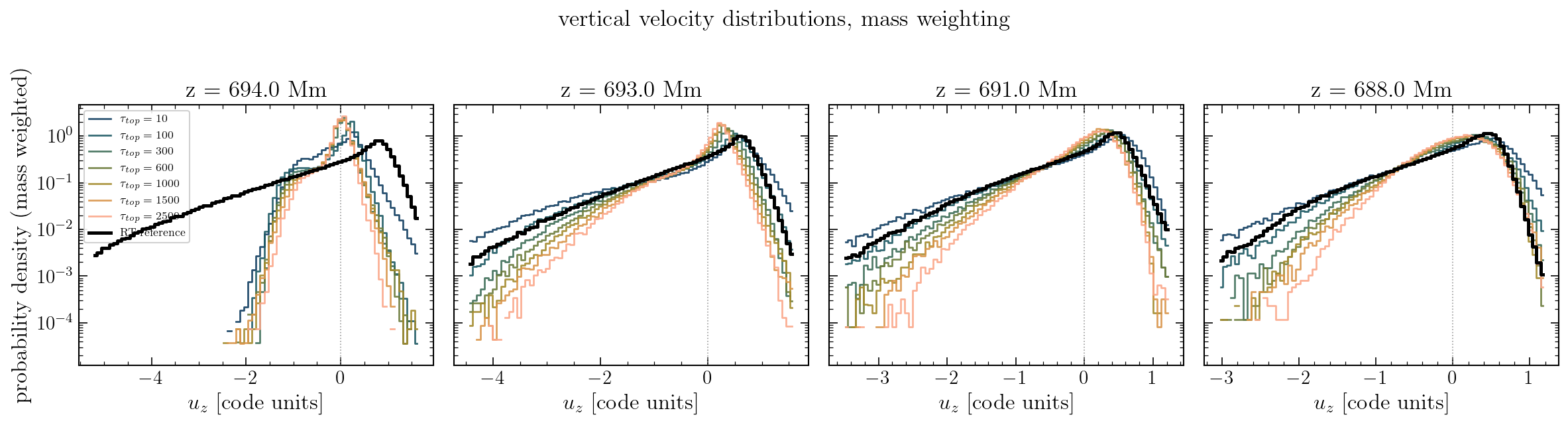}}
  \caption{Vertical velocity distributions, used to calibrate $\tau_{\rm top}$.}
  \label{fig:cal_uz_hist}
\end{figure*}

The second diagnostic is the distribution of vertical velocity at several depths (Figure \ref{fig:cal_uz_hist}), which responds to the strength of the cooling. Here we want the velocity distribution to be as close to the reference simulation as possible. Runs with too long timescale produce insufficient cold material and too weak a downflow tail; runs with too short $\tau_{\rm top}$ overshoot, producing a cold uniform `blanket` on top and the velocity distribution becomes too broad, with peak position shifted. 

\begin{figure*}
  \centering
  \resizebox{\hsize}{!}
  {\includegraphics[width=0.99\columnwidth]{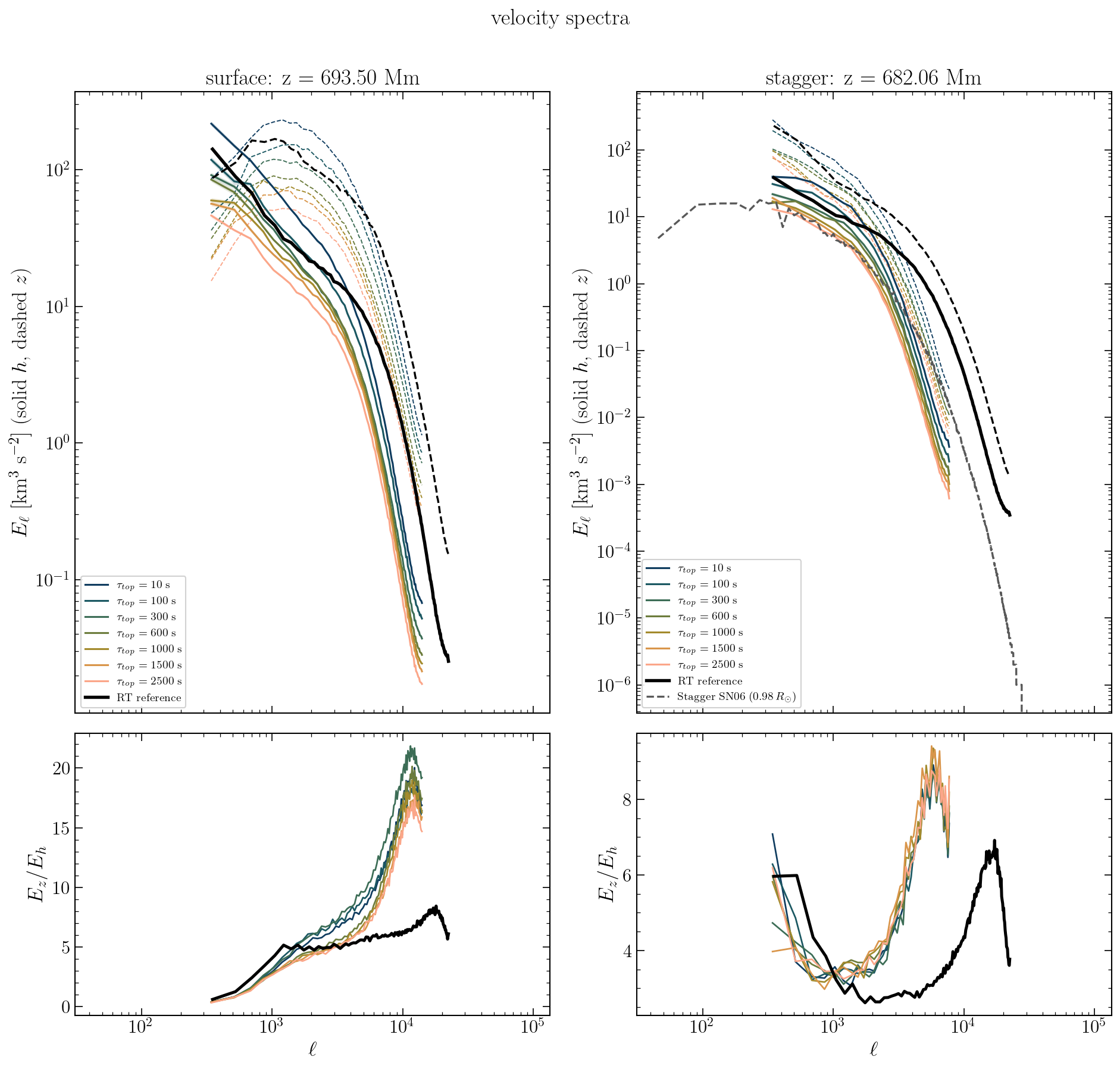}}
  \caption{Horizontal and vertical velocity power spectra, used to calibrate $\tau_{\rm top}$.}
  \label{fig:cal_power}
\end{figure*}

The third diagnostic is the velocity power spectrum, compared with the reference, shown in Figure \ref{fig:cal_power}. We compute both the horizontal and the vertical spectra, since they constrain different things. The granule size distribution is carried by $E_h$, while $E_z$ reflects the strength of the driving; a mismatch appearing in $E_h$ alone identifies a granule-scale error, whereas one appearing in both identifies a driving-strength error. The ratio $E_z/E_h$, being dimensionless, is insensitive to the overall normalisation and is reported alongside.

We use two reference models: (1) The radiative-transfer reference constrains the amplitude and its depth dependence; (2) The power spectrum at $0.98\,R_\odot$ from Stagger simulation \citep{Nordlund_2009} constrains the shape --- the position of the spectral peak and the slopes, but not necessarily the absolute level, since that simulation has its own upper boundary conditions and different resolution. Spectra are therefore evaluated both just below the truncation radius, where the granule-splitting signature is strongest, and at $0.98\,R_\odot$, where the external comparison is available. No value of $\tau_{\rm top}$ reproduces $E_z/E_h$ at granular scales. Just below $r_{\rm top}$ the ratio peaks at $\approx$ 20, against $\approx$ 7 in the RT reference, independently of $\tau_{\rm top}$. The parametrized cooling therefore over-drives vertical relative to horizontal motions at these scales. 

Since parametric surface cooling at certain depths below the optical surface can't exactly replicate the full radiative transfer, an exact match is not possible. The parametrized surface cooling should thus be regarded as a trade-off, which makes cheaper simulations possible but results in less authoritative results. In the future we will explore the possibility to calibrate and employ entropy rain `patches' \citep{Kapyla_2025} for a better cooling prescription.

Given the calibration of parameters, the configuration was tested with an extended duration run, to evaluate whether the emergent flux remains at an acceptable level.

\section{Large scale simulations} \label{sec:photosphere}

\begin{table}
\caption{Configuration of the whole convection zone experiment.}            
\label{tab:ph_sim_setup}      
\centering                         
\begin{tabular}{l c}        
\hline                       
   Bottom radius & 0.655 R$_\odot$ \\      
   Top radius & 1.0007 R$_\odot$ \\
   N$_\theta$ & 260 \\
   N$_{\theta, min}$ & 36 \\
   N$_{\theta}$ reduction rate & 80, 60, 40, 8, 4, 4, ... \\
   $\Delta s_{min}$ & 71 km \\
   EOS & FreeEOS \citep{Irwin_2012} \\
   Opacities & Blue package \citep{Amarsi_2016} \\
   Radiative transfer & 4 bin, 1 ray dir \\
   MHD solver & HLLS \citep{Popovas_2025a} \\
   Time unit & $10^3$ s \\
   Length unit & $10^6$ m (1 Mm) \\
   Mass unit & $10^{23}$ g \\
   $\Omega_{coriolis}$ & 424 nHz \\ 
\hline                                  
\end{tabular}
\end{table}

\begin{figure*}
  \centering
  \resizebox{\hsize}{!}
  {\includegraphics[width=0.99\columnwidth]{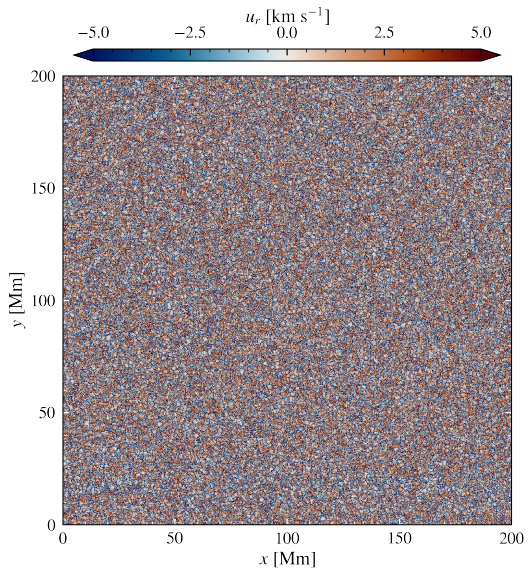}}
  \caption{Radial velocity cutout in km/s at $R=R_\odot$ (500 km below the upper BC) in a test simulation of the entire convection zone of the Sun.}
  \label{fig:vb_rt_surf}
\end{figure*}

The Table \ref{tab:ph_sim_setup} summarises the main parameters for the large scale experiment setup. The domain spans from 0.655 R$_\odot$ to ~500 km above the surface, thus comfortably covers the entire convection zone. For surface cooling we have used the multi-bin short characteristic radiative transfer (only purely radial rays though) using the Blue opacity package \citep{Amarsi_2016}. The smallest cell size near the surface was 71 km. The simulation was run on LUMI using 504 nodes. A radial velocity cutout at $R=R_\odot$ is shown in Figure \ref{fig:vb_rt_surf}. Typical granule size is ~1--1.5 Mm, typical radial velocities ~1-3 km/s, both in the expected range of the observed solar surface. However, this simulation exemplified one of the main challenges - each snapshot was 2.3 terabyte large and analysing such datasets becomes as expensive and complicated as running the simulation. Visualisation of the entire surface is also extremely challenging and underwhelming - granules are too small to be visible in a figure without excessive zooming.

\section{Weak and strong scaling} \label{strong_weak_scaling}
The DISPATCH framework has demonstrated excellent scaling results past 150,000 cores on a fluid dynamics problem \citep{Nordlund_2018MNRAS}. These results stem from the nearest-neighbour-only MPI communications, combined with the near-perfect scaling of OpenMP-parallel task updates on each rank, and thus do not depend on the detailed properties of the solver. Non-Cartesian geometry implies slightly less straightforward MPI decomposition, which, in turn might affect load balancing. In volleyball seams patches can have more than 26 neighbours with more awkward overlapping regions of authority. All this, together with additional (experiment-specific) physics, such as gravity, Coriolis force, etc., could affect the perfect scaling that was previously demonstrated in simpler geometries \citep{Nordlund_2018MNRAS}. Thus we conducted new weak and strong scaling tests with the current experiment, to check that excellent scalability can still be achieved. The tests were conducted on the LUMI and Betzy supercomputer systems. Both of these machines have AMD EPYC CPUs (models 7763 and 7742 respectively), which makes them very similar, to the extent that Betzy can be considered LUMI's `little sister'.  The two machines have different node interconnect and different compilers that we use, though, Cray/GCC and GCC respectively. Thus, the performance differences are due to the differences in the system configurations and software.

\begin{figure}
\centering
    \includegraphics[width=0.85\columnwidth]{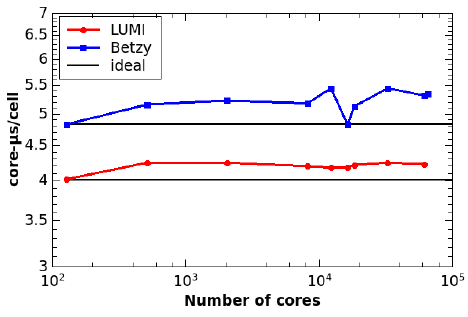}
    \caption{Weak scaling results on LUMI (red) and Betzy (blue) supercomputers. For both machines the ideal scaling is shown in black.}
    \label{fig:weak_scaling}
\end{figure}

To test the weak scaling, we employ our simplified 'sandbox' runs, where we fix the vertical resolution and expand in the horizontal plane with an increasing number of cores. We fix the patch dimensions to $24^3$ cells and keep the number of patches per compute node constant. In all the runs there are 2 MPI ranks per node. In this test we take the average update time (in $\rm{\mu}$s/cell) over an extended period of run-time in the execution phase (we ignore the initialization phase). \Figure{weak_scaling} shows that with our experimental setup we can still achieve nearly perfect weak scaling.

\begin{figure}
\centering
    \includegraphics[width=0.85\columnwidth]{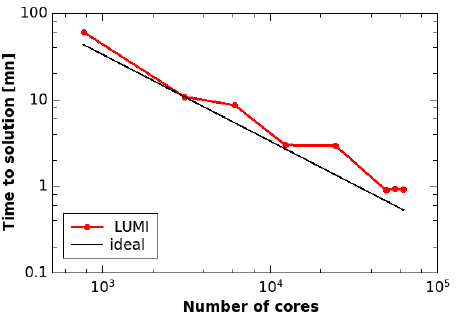}
    \caption{Strong scaling results on LUMI supercomputer.}
    \label{fig:strong_scaling}
\end{figure}

To test the strong scaling we run the complete experimental setup. This setup is kept fixed to check that different MPI decomposition (albeit still carefully chosen to have reasonably good initial load balance) does not negatively affect the run-time and the experiment can be successfully executed on a range of MPI decomposition configurations. In this test we measure the total wall-clock time it takes for an experiment to get from the start of the execution until experiment time reaches 0.2 code units. With a typical time-step of the order of approximately 10$^{-3}$--10$^{-2}$, each patch is updated between of order 20 to 200 times. \Figure{strong_scaling} shows the result and it is clear that the volleyball mesh decomposition together with the experimental setup has very good strong scaling properties. Note, that at the highest number of cores the efficiency drops, as the number of patches drops to about 2.5 patches per core, which is well below the optimal 20-50 tasks per core. The nearly perfect scaling is re-established if mesh-refinement is used and the workload (the number of patches per core) increases.

\end{appendix}
\end{document}